1  **Strong Ground Motion in the 2011 Tohoku Earthquake:**

2  **a 1Directional - 3Component Modeling**

4  by Maria Paola Santisi d'Avila, Jean-François Semblat and Luca Lenti

11  Corresponding author:

12  Maria Paola Santisi d'Avila

13  University of Nice Sophia Antipolis - Laboratoire Jean Alexandre Dieudonné

15  Address:

16  14bis, Rue François Guisol - 06300 Nice - France

17  Phone: +33(0)4 92 07 69 96

18  Email: msantisi@unice.fr

19       mpaolasantisi@gmail.com

**ABSTRACT**

Local wave amplification due to strong seismic motions in surficial multilayered soil is influenced by several parameters such as the wavefield polarization and the dynamic properties and impedance contrast between soil layers. The present research aims at investigating seismic motion amplification in the 2011 Tohoku earthquake through a one-directional three-component (1D-3C) wave propagation model. A 3D nonlinear constitutive relation for dry soils under cyclic loading is implemented in a quadratic line finite element model. The soil rheology is modeled by mean of a multi-surface cyclic plasticity model of the Masing-Prandtl-Ishlinskii-Iwan (MPII) type. Its major advantage is that the rheology is characterized by few commonly measured parameters. Ground motions are computed at the surface of soil profiles in the Tohoku area (Japan) by propagating 3C signals recorded at rock outcrops, during the 2011 Tohoku earthquake. Computed surface ground motions are compared to the Tohoku earthquake records at alluvial sites and the reliability of the 1D-3C model is corroborated. The 1D-3C approach is compared with the combination of three separate one-directional analyses of one motion component propagated independently (1D-1C approach). The 3D loading path due to the 3C-polarization leads to multiaxial stress interaction that reduces soil strength and increases nonlinear effects. Time histories and spectral amplitudes, for the Tohoku earthquake, are numerically reproduced. The 1D-3C approach allows the evaluation of various parameters of the 3C motion and 3D stress and strain evolution all over the soil profile.

**INTRODUCTION**

One-directional wave propagation analyses are an easy way to estimate the surface ground motion, even in the case of strong seismic events. Seismic waves due to strong ground motions propagating in surficial soil layers may both reduce soil stiffness and increase nonlinear effects. The nonlinear behavior of the soil may have beneficial or detrimental effects on the dynamic response at the surface, depending on the energy dissipation rate. The three-dimensional (3D) loading path also influences the stresses into the soil and thus its seismic response.

The recent records of the 9 Mw 11 March 2011 Tohoku earthquake, in Japan, allow to understand the influence of incident wave polarization. This event is one of the largest earthquakes in the world that has been well recorded in the near-fault zone. According to the Japanese database of the K-Net accelerometer network (see Data and Resources Section), the main feature of the Tohoku three-component records is that the vertical to maximum horizontal component ratio appears close to one for several soil profiles and the peak vertical motion can locally be higher than the minor horizontal component of ground motion. This is an interesting observation because earthquake vertical component was neglected in structural design codes in the recent past. The vertical to horizontal ratio, previously considered trivial, becomes essential to characterize 3D loading effects and multiaxial stress interaction in strong ground motion modeling.

In order to investigate site-specific seismic hazard, past studies have been devoted to one-directional shear wave propagation in a multilayered soil profile (1D-propagation) considering one motion component only (1C-polarization). Several one-directional models and related codes were developed, to investigate one-component ground response of horizontally layered sites, reproducing soil behavior as equivalent linear (SHAKE, Schnabel *et al*., 1972; EERA, Bardet *et al*., 2000), dry nonlinear (NERA, Bardet *et al*., 2001) and saturated nonlinear (DESRA-2, Lee and Finn, 1978).

Soils are complex materials and a linear approach is not reliable to model their seismic response to strong earthquakes. The continuous improvement of dynamic test apparatus allows to measure dynamic soil properties over a wide range of strains, showing the highly nonlinear deformation characteristics of soil and the significant variation of shear modulus and damping ratio with the amplitude of shear strain under cyclic loading (Seed and Idriss, 1970a; Hardin and Drnevich, 1972a, 1972b; Kim and Novak, 1981; Lefebvre *et al*., 1989; Vucetic and Dobry, 1991; Vucetic, 1994; Ishihara, 1996; Hsu and Vucetic, 2004, 2006). At larger strain levels, the nonlinearity may reduce the shear modulus and increase the damping. Observations in situ enabled to undertake quantitative studies on the nonlinear response of soft sedimentary sites and to evaluate local site effects (Seed and Idriss, 1970b; Satoh *et al.*, 1995; Bonilla *et al*., 2002; De Martin *et al*., 2010).

A nonlinear site response analysis accounting for hysteresis allows to follow the time evolution of the stress and strain during seismic events and to estimate the resulting surface seismic ground motion at large strain levels. The nonlinear analysis requires the propagation of a seismic wave in a nonlinear medium by using an appropriate constitutive model and integrating the wave equation in the time domain. Inputs to these analyses include acceleration time histories at bedrock and nonlinear material properties of the various soil strata underlying the site. The main difficulty in nonlinear analysis is to find a constitutive model that reproduces faithfully the nonlinear and hysteretic behavior of soil under cyclic loadings, with the minimum number of parameters.

Considering the 3D loading path means representing the 3D hysteretic behavior of soils, which is difficult to model because the yield surface may present a complex form. The nonlinear 3D constitutive behavior depends on the 3D loading path. The three motion components are coupled, due to the nonlinear 3D constitutive behavior, and they cannot be computed separately (Li *et al*., 1992; Santisi d'Avila *et al*., 2012). Li (1990) incorporated the three-dimensional cyclic plasticity soil model proposed by Wang *et al*. (1990) in a 1D finite element procedure (SUMDES code, Li *et al*., 1992), in terms of effective stress, to simulate the one-directional wave propagation accounting for pore pressure in the soil. However, this complex rheology needs an excessive number of parameters to characterize the soil model.

In the present research, the nonlinear soil behavior is represented by the so-called Masing-Prandtl-Ishlinskii-Iwan (MPII) model, according to (Segalman and Starr, 2008), or Iwan's model (Iwan, 1967). It is a multi-surface plasticity mechanism for cyclic loading and it depends on few parameters that can be obtained from ordinary laboratory tests. Material properties include the dynamic shear modulus at low strain and the variation of shear modulus with shear strain. This rheology allows the dry soil to develop large strains in the range of stable nonlinearity, where the shape of hysteresis loops remains unvaried in the time. Due to its three-directional nature, the procedure can handle both shear wave and compression wave simultaneously and predict not only horizontal motion but vertical settlement too.

The implementation of the MPII nonlinear cyclic constitutive model in a finite element scheme (SWAP_3C code) is presented in detail by Santisi d'Avila *et al.* (2012). The authors analyze the importance of a three-directional shaking problem, evaluating the seismic ground motion due to three-component strong earthquakes, for well-known stratigraphies, using synthetic incident wavelets. The role of critical parameters affecting the soil response is investigated. The main feature of the procedure is that it solves the specific three-dimensional stress-strain problem for seismic wave propagation along one-direction only, using a constitutive behavior depending only on commonly measured soil properties.

In the present research, the goal is to assess the reliability of the model proposed by the authors

(Santisi d'Avila *et al.*, 2012) and confirm, through actual data, the findings of the parametric analysis previously done using synthetic wavelets. It was observed that the shear modulus decreases and the dissipation increases, for a given maximum strain amplitude, from one to three component unidirectional propagated wave. The material strength is lower under triaxial loading rather than for simple shear loading. The shape of hysteresis loops remains unvaried in the time, for one-component loading, in the strain range of stable nonlinearity. In the case of three-component loading, the shape of the hysteresis loops changes in the time for shear strains in the same range. Hysteresis loops for each horizontal direction are altered as a consequence of the interaction between loading components. The main difference between three superimposed one-component ground motions (1D-1C approach) and the proposed one-directional three-component propagation model (1D-3C approach) is remarkable in terms of ground motion time history, maximum stress and hysteretic behavior, with more nonlinearity and coupling effects between components. This kind of consequence is more evident with decreasing seismic velocity ratio in the soil and increasing vertical to horizontal component ratio of the incident wave.

The 1D-3C propagation model and the main features of the applied constitutive relation are presented. The validation of the 1D-3C approach is undertaken comparing the three-component records of the 2011 Tohoku earthquake with numerical time histories. Seismic records with vertical to horizontal acceleration ratio higher than 70 % are applied to investigate the impact of a large vertical to horizontal peak acceleration ratio. The simultaneous propagation of a three component input signal, in a system of horizontal soil layers, is studied using the proposed model. The case of three components simultaneously propagated (1D-3C) is compared with that of three superimposed one-component ground motions (1D-1C), to understand the influence of a

139  3D loading path and input wavefield polarization. The influence of the soil properties and quake
140  features on the local seismic response is discussed for the case of multilayered soil profiles in
141  the Tohoku area (Japan).

142

143  **ONE-DIRECTIONAL THREE-COMPONENT PROPAGATION MODEL**

144  The three components of the seismic motion are propagated along one direction in a nonlinear
145  soil profile from the top of the underlying elastic bedrock. The multilayered soil is assumed
146  infinitely extended along the horizontal directions. Shear and pressure waves propagate
147  vertically in the $z$-direction. These hypotheses yield no strain variation in $x$- and $y$-direction.
148  At a given depth, soil is assumed to be a continuous and homogeneous medium.
149  Transformations remain small during the process and the cross sections of three-dimensional
150  soil elements remain planes.

151

152  **Spatial discretization**

153  Soil stratification is discretized into a system of horizontal layers, parallel to the $xy$ plane, by
154  using a finite element scheme (Fig. 1). Quadratic line elements with three nodes are considered.
155  According to the finite element modeling, the discrete form of equilibrium equations, is
156  expressed in the matrix form as

157  $$\mathbf{M}\ddot{\mathbf{D}} + \mathbf{C}\dot{\mathbf{D}} + \mathbf{F}_{int} = \mathbf{F} \qquad (1)$$

158  where $\mathbf{M}$ is the mass matrix, $\dot{\mathbf{D}}$ and $\ddot{\mathbf{D}}$ are velocity and acceleration vectors, respectively, i.e.
159  the first and second time derivatives of the displacement vector $\mathbf{D}$. $\mathbf{F}_{int}$ is the vector of nodal
160  internal forces and $\mathbf{F}$ is the load vector. $\mathbf{C}$ is a damping matrix derived from the chosen
161  absorbing boundary condition. The differential equilibrium problem (1) is solved according to

162 compatibility conditions and the hypothesis of no strain variation in the horizontal directions, to
163 a three-dimensional nonlinear constitutive relation for cyclic loading and the boundary
164 conditions described below.

165 Discretizing the soil column into $n_e$ quadratic line elements and consequently into $n = 2n_e + 1$
166 nodes (Fig. 1), having three translational degrees of freedom each, yields a $3n$-dimensional
167 displacement vector **D** composed by three blocks whose terms are the displacement of the $n$
168 nodes in $x$-, $y$- and $z$-direction, respectively. Soil properties are assumed constant in each
169 finite element and soil layer.

170 The minimum number of quadratic line elements per layer $n_e^j$ is defined considering that $p = 10$
171 is the minimum number of nodes per wavelength to accurately represent the seismic signal
172 (Kuhlemeyer and Lysmer, 1973; Semblat and Brioist, 2000) and it is evaluated as

$$\min n_e^j = \frac{H_j}{2} \frac{p f}{v_s} \tag{2}$$

174 where $H_j$ is the thickness of layer $j$ (Fig. 1), $f$ is the assumed maximum frequency of the
175 input signal and $v_s$ is the assumed minimum shear velocity in the medium. The seismic signal
176 wavelength is equal to $v_s/f$. The assumed minimum $v_s$ is related to the assumed maximum
177 shear modulus decay and allows to account for non linearities. In this study, $v_s$ corresponds to a
178 70% reduction of the initial shear modulus. The maximum frequency $f$, used to assess the
179 minimum number of elements per layer $n_e^j$, is assumed to be 15 Hz as an accurate choice.

180 The assemblage of $(3n \times 3n)$-dimensional matrices and $3n$-dimensional vectors is independently
181 done for each of the three $(n \times n)$-dimensional submatrices and $n$-dimensional subvectors,
182 respectively, corresponding to $x$-, $y$- and $z$-direction of motion.

183 **Boundary conditions**

184 The system of horizontal soil layers is bounded at the top by the free surface and at the bottom
185 by a semi-infinite elastic medium representing the seismic bedrock. The stresses normal to the
186 free surface are assumed null and the following condition, implemented by Joyner and Chen
187 (1975) and Joyner (1975) in a finite difference formulation and used by Bardet and Tobita
188 (2001) in NERA code, is applied at the soil-bedrock interface to take into account the finite
189 rigidity of the bedrock:

190 $$-\mathbf{p}^T \boldsymbol{\sigma} = \mathbf{c}(\mathbf{v} - 2\mathbf{v}_b) \qquad (3)$$

191 The stresses normal to the soil column base at the bedrock interface are $\mathbf{p}^T \boldsymbol{\sigma}$ and $\mathbf{c}$ is a $(3\times 3)$
192 diagonal matrix whose terms are $\rho_b v_{sb}$, $\rho_b v_{sb}$ and $\rho_b v_{pb}$. The parameters $\rho_b$, $v_{sb}$ and $v_{pb}$ are
193 the bedrock density and shear and pressure wave velocities in the bedrock, respectively. The
194 three terms of vector $\mathbf{v}$ are the unknown velocities in $x$-, $y$- and $z$-direction, respectively, at
195 the interface soil-bedrock (node 1 in Fig. 1). The terms of the 3-dimensional vector $\mathbf{v}_b$ are the
196 input bedrock velocities, in the underlying elastic medium in directions $x$, $y$ and $z$,
197 respectively. Boundary condition (3) allows energy to be radiated back into the underlying
198 medium.

199 The three-component bedrock velocity can be obtained by halving seismic records at
200 outcropping bedrock. The incident bedrock waves are the half of outcropping seismic waves
201 (Fig. 1), due to the free surface effect in linear elastic medium such as rock.

202 If borehole records are used, the halving operation is not necessary, because records are applied
203 as incident bedrock signals. The bedrock is assumed elastic in the proposed model, with
204 absorption and reflection of waves at the soil-bedrock interface, according to equation (3).
205 However, the borehole input signal contains incident and reflected waves. The absorbing condition in equation (3) is commonly used also when borehole records are applied (NERA code, Bardet and Tobita, 2001), but an imposed motion at the soil-bedrock interface (first node) would more properly represent the borehole boundary condition. The implementation of such a boundary condition, adopted when borehole records are analyzed, will be a future improvement of the proposed procedure.

**Time discretization**

The finite element model and the soil nonlinearity require spatial and time discretization, respectively, to permit the problem solution (Hughes, 1987; Crisfield, 1991). The rate type constitutive relation between stress and strain is linearized at each time step. Accordingly, equation (1) is expressed as

$$\mathbf{M}\,\Delta\ddot{\mathbf{D}}_k^i + \mathbf{C}\,\Delta\dot{\mathbf{D}}_k^i + \mathbf{K}_k^i\,\Delta\mathbf{D}_k^i = \Delta\mathbf{F}_k \qquad (4)$$

where the subscript $k$ indicates the time step $t_k$ and $i$ the iteration of the problem solving process, as explained below.

The step-by-step process is solved by the Newmark algorithm, expressed as follows:

$$\begin{cases} \Delta\dot{\mathbf{D}}_k^i = \dfrac{\gamma}{\beta\,\Delta t}\Delta\mathbf{D}_k^i - \dfrac{\gamma}{\beta}\dot{\mathbf{D}}_{k-1} + \left(1 - \dfrac{\gamma}{2\beta}\right)\Delta t\,\ddot{\mathbf{D}}_{k-1} \\ \Delta\ddot{\mathbf{D}}_k^i = \dfrac{1}{\beta\,\Delta t^2}\Delta\mathbf{D}_k^i - \dfrac{1}{\beta\,\Delta t}\dot{\mathbf{D}}_{k-1} - \dfrac{1}{2\beta}\ddot{\mathbf{D}}_{k-1} \end{cases} \qquad (5)$$

The Newmark's procedure is an implicit self-starting unconditionally stable approach for one-step time integration in dynamic problems (Newmark, 1959; Hilber *et al.*, 1977; Hughes, 1987). The two parameters $\beta = 0.3025$ and $\gamma = 0.6$ guarantee unconditional stability of the time integration scheme and numerical damping properties to damp higher modes (Hughes, 1987). Equations (4) and (5) yield

227 $$\bar{\mathbf{K}}_k^i \Delta \mathbf{D}_k^i = \Delta \mathbf{F}_k + \mathbf{A}_{k-1} \tag{6}$$

228 where the modified stiffness matrix is defined as

229 $$\bar{\mathbf{K}}_k^i = \frac{1}{\beta \Delta t^2}\mathbf{M} + \frac{\gamma}{\beta \Delta t}\mathbf{C} + \mathbf{K}_k^i \tag{7}$$

230 and $\mathbf{A}_{k-1}$ is a vector depending on the response in previous time step, given by

231 $$\mathbf{A}_{k-1} = \left[\frac{1}{\beta \Delta t}\mathbf{M} + \frac{\gamma}{\beta}\mathbf{C}\right]\dot{\mathbf{D}}_{k-1} + \left[\frac{1}{2\beta}\mathbf{M} + \left(\frac{\gamma}{2\beta}-1\right)\Delta t\, \mathbf{C}\right]\ddot{\mathbf{D}}_{k-1} \tag{8}$$

232 Equation (4) requires an iterative solving, at each time step $k$, to correct the tangent stiffness

233 matrix $\mathbf{K}_k^i$. Starting from the stiffness matrix $\mathbf{K}_k^1 = \mathbf{K}_{k-1}$, evaluated at the previous time step, the

234 value of matrix $\mathbf{K}_k^i$ is updated at each iteration $i$ (Crisfield, 1991). After evaluating the

235 displacement increment $\Delta \mathbf{D}_k^i$ by equation (6), using the tangent stiffness matrix corresponding to

236 the previous time step, velocity and acceleration increments can be estimated through equation

237 (5) and the total motion is obtained according to

238 $$\mathbf{D}_k^i = \mathbf{D}_{k-1} + \Delta \mathbf{D}_k^i \quad \dot{\mathbf{D}}_k^i = \dot{\mathbf{D}}_{k-1} + \Delta \dot{\mathbf{D}}_k^i \quad \ddot{\mathbf{D}}_k^i = \ddot{\mathbf{D}}_{k-1} + \Delta \ddot{\mathbf{D}}_k^i \tag{9}$$

239 where $\mathbf{D}_k^i$, $\dot{\mathbf{D}}_k^i$ and $\ddot{\mathbf{D}}_k^i$ are the vectors of total displacement, velocity and acceleration,

240 respectively. The strain increments are then derived from the displacement increments, terms of

241 vector $\Delta \mathbf{D}_k^i$. Stress increments and tangent constitutive matrix are obtained through the assumed

242 constitutive relationship. Gravity load is imposed as static initial condition in terms of strain and

243 stress at nodes. The modified stiffness matrix $\bar{\mathbf{K}}_k^i$ is calculated and the process restarts. The

244 correction process continues until the difference between two successive approximations is

245 reduced to a fixed tolerance, according to

246 $$\left|\mathbf{D}_k^i - \mathbf{D}_k^{i-1}\right| < \alpha \left|\mathbf{D}_k^i\right| \tag{10}$$

247  where $\alpha = 10^{-3}$ (Mestat, 1993, 1998). Afterwards, the next time step is analyzed.

248

249  **FEATURES OF THE 3D NONLINEAR HYSTERETIC MODEL**

250  The three-dimensional constitutive model for soil used to model the propagation of a three-
251  component earthquake, in stratified soils, is a Masing-Prandtl-Ishlinskii-Iwan (MPII) type
252  constitutive model (Segalman and Starr, 2008), suggested by Iwan (1967) and applied by Joyner
253  (1975) and Joyner and Chen (1975) in a finite difference formulation. It is used in the present
254  work to properly model the nonlinear soil behavior in a finite element scheme (Santisi d'Avila *et
255  al.*, 2012).

256  The so-called Masing rules, presented in 1926, describe the loading and unloading paths in the
257  stress-strain space, reproducing quite faithfully the hysteresis observed in the laboratory. Prandtl
258  proposed, in 1928, an elasto-plastic model with strain-hardening, re-examined by Ishlinskii in
259  1944, obtained by coupling a family of stops in parallel or of plays in series (Bertotti and
260  Mayergoyz, 2006). Iwan (1967) proposed an extension of the standard incremental theory of
261  plasticity (Fung, 1965), by introducing a family of yield surfaces, modifying the 1D approach
262  with a single yield surface in the stress space. He modeled nonlinear stress-strain curves using a
263  series of mechanical elements, having decreasing stiffnesses and increasing sliding resistance.
264  The MPII model takes into account the nonlinear hysteretic behavior of soils in a three-
265  dimensional stress state, using an elasto-plastic approach with hardening, based on the definition
266  of a series of nested yield surfaces, according to von Mises' criterion. The MPII model is used to
267  represent the behavior of materials satisfying Masing criterion (Kramer, 1996) and not
268  depending on the number of loading cycles. The stress level depends on the strain increment and
269  strain history but not on the strain rate. Therefore, this rheological model has no viscous damping

270  and the energy dissipation process is purely hysteretic and does not depend on the frequency.

271  Shear modulus and damping ratio are strain-dependent.

272  The main feature of the MPII rheological model is that the only necessary input data, to identify

273  soil properties in the applied constitutive model, is the shear modulus decay curve $G(\gamma)$ versus

274  shear strain $\gamma$. The initial elastic shear modulus $G_0 = \rho v_s^2$, measured at the elastic behavior range

275  limit $\gamma \cong 0.001‰$ (*Fahey*, 1992), depends on the mass density $\rho$ and the shear wave velocity in

276  the medium $v_s$. The P-wave modulus $M = \rho v_p^2$, depending on the pressure wave velocity in the

277  medium $v_p$, characterizes the longitudinal behavior of soil. The seismic velocity ratio

278  (compressional to shear wave velocity ratio $v_p/v_s$), evaluated by

279  $$\left(v_p/v_s\right)^2 = 2(1-\nu)/(1-2\nu) \qquad (11)$$

280  is a function of the Poisson's ratio $\nu$. This is a parameter of the constitutive behavior for

281  multiaxial load and of the interaction between components in the three-dimensional response.

282  The MPII hysteretic model for dry soils, used in the present research, is applied for strains in the

283  range of stable nonlinearity. In this range, where the shear strain is lower than the stability

284  threshold (Lefebvre *et al.*, 1989), both shear modulus and damping ratio do not depend on the

285  number of cycles. Stable stress-strain cycles are observed, for which the shape of hysteresis

286  loops remains unvaried at each cycle, for one-component loading. When the stability threshold is

287  overtaken, the soil mechanical response changes at each cycle and both shear modulus and

288  damping ratio vary abruptly (Zambelli *et al.*, 2006). Unstable liquefaction phenomena appear for

289  large shear strains and, consequently, both the hysteresis loop shape and the average shear

290  stiffness evolve progressively with the number of cycles.

291  Large strain rates are not adequately reproduced without taking into account undrained condition

292 for soils. Constitutive behavior models for saturated soils would allow to attain larger strains
293 with proper accuracy. It is the reason why the shear modulus decay is accepted until 70 %,
294 corresponding to the minimum shear velocity in the soil in equation (2), used to obtain an
295 appropriate space discretization.

296 In the present study the soil behavior is assumed adequately described by a hyperbolic stress-
297 strain curve (Hardin and Drnevich, 1972b). This assumption yields a normalized shear modulus
298 decay curve, used as input curve representing soil characteristics, expressed as

299 $$G/G_0 = 1/\left(1+\left|\gamma/\gamma_r\right|\right) \quad (12)$$

300 where $\gamma_r$ is a reference shear strain corresponding to an actual tangent shear modulus equivalent
301 to 50 % of the initial shear modulus, in a normalized shear modulus decay curve provided by
302 laboratory test data. The applied constitutive model (Iwan, 1967; Joyner and Chen, 1975; Joyner,
303 1975) does not depend on the hyperbolic initial loading curve. It could incorporate also shear
304 modulus decay curves obtained from laboratory dynamic tests on soil samples.

305 The stiffness matrix $\mathbf{K}_k^i$ is deduced, at each time step $k$ and iteration $i$, knowing the tangent
306 constitutive matrix $\mathbf{E}_k^i$. The actual strain level and the strain and stress values at the previous
307 time step allow to evaluate the tangent constitutive (6x6) matrix $\mathbf{E}_k^i$ and the stress increment,
308 according to the incremental constitutive relationship $\Delta\boldsymbol{\sigma}_k^i = \mathbf{E}_k^i \Delta\boldsymbol{\varepsilon}_k^i$. The deviatoric constitutive
309 matrix $\mathbf{E}_d$ for a three-dimensional soil element is obtained according to Iwan's procedure, as
310 presented by Joyner (1975), and allows to evaluate the vector of deviatoric stress increments $\Delta\mathbf{s}$,
311 knowing the vector of deviatoric strain increments $\Delta\mathbf{e}$, according to $\Delta\mathbf{s} = \mathbf{E}_d \Delta\mathbf{e}$. The total
312 constitutive matrix $\mathbf{E}$ is evaluated starting from $\mathbf{E}_d$ (Santisi d'Avila *et al.*, 2012).

313 Stress and strain rate in the one-dimensional (1D) soil profile due to the propagation of a three-

314 component earthquake are expressed in the following analysis in terms of octahedral shear stress
315 and strain, accounting for the hypothesis of infinite horizontal soil $\left(\varepsilon_{xx}=0, \varepsilon_{yy}=0, \gamma_{xy}=0\right)$.
316 According to the 3D constitutive model and for null $\gamma_{xy}$, the only null stress component is the
317 in-plane shear stress $\tau_{xy}$. Octahedral stress (respectively strain) is chosen to combine the three-
318 dimensional stress (respectively strain) components in a unique scalar parameter, that allows an
319 adequate comparison of the simultaneous propagation of the three motion components (1D-3C)
320 and the independent propagation of the three components (1D-1C) superposed a posteriori. The
321 1D-1C approach is a good approximation in the case of low strains within the linear range
322 (superposition principle, Oppenheim *et al.*, 1997). The effects of axial-shear stress interaction in
323 multiaxial stress states have to be taken into account for higher strain rates, in the nonlinear
324 range. The octahedral stress and strain are respectively obtained by

$$
\begin{aligned}
\tau_{oct} &= \frac{1}{3}\sqrt{\left(\sigma_{xx}-\sigma_{yy}\right)^2+\left(\sigma_{yy}-\sigma_{zz}\right)^2+\left(\sigma_{zz}-\sigma_{xx}\right)^2+6\left(\tau_{yz}^2+\tau_{zx}^2\right)} \\
\gamma_{oct} &= \frac{2}{3}\sqrt{2\left(\varepsilon_{zz}\right)^2+6\left(\varepsilon_{yz}^2+\varepsilon_{zx}^2\right)}
\end{aligned}
\tag{13}
$$

326

327 **VALIDATION OF THE 1D-3C WAVE PROPAGATION MODELING**
328 Recorded data from the 9 Mw 11 March 2011 Tohoku earthquake by the K-Net and KiK-Net
329 accelerometer networks have been analyzed in this research (see Data and Resources Section), to
330 numerically reproduce the surface ground motion and to provide non-measured parameters.
331 Kyoshin Network (K-Net) database stores ground motion records at the surface of soil profiles
332 and related stratigraphies; whereas, the Kiban-Kyoshin Network (KiK-Net) database provides
333 surface and borehole seismic records for different stratigraphies.
334 We use records at the surface of alluvial soil profiles to validate the numerical surface ground motion computed by the proposed model. Some rock type profiles close to each analyzed soil profile are selected (Fig. 2), in the K-Net database (see Data and Resources Section), to get incident seismic motion at the base of the profiles. Incident seismic motion at the base of soil profiles is the halved motion at a close outcropping bedrock site (Fig. 1). Incident and surface seismic motions are known in the case of KiK-Net stratigraphies, according to the assumption that borehole signals are applied as incident. As explained before, this improper adoption will be overcome in a later work.

The numerical one-directional dynamic response of studied soil profiles is validated by comparison with recordings in terms of acceleration time histories at the ground surface, since it is the only available recorded data. The numerical acceleration time history is obtained by the estimated velocity time history after derivation and low-pass filtering (to $10\,\text{Hz}$). The three-component ground motion is characterized by the modulus which is a unique scalar parameter. Spectral amplitudes are compared and discussed below.

**Soil profiles**

The soil columns modeled in this study, consisting of various layers on seismic bedrock, are analyzed to validate the 1D-3C wave propagation modeling by using real data and to investigate the local seismic response by the 1D-3C approach. The stratigraphic setting of four soil profiles in the Tohoku area (Japan) is used in this analysis (Table 1). The description of the stratigraphy and lithology of the alluvial deposits in the Tohoku area is provided by the Kyoshin Network database (see Data and Resources Section). Average shear wave velocities and epicentral distances are listed in Table 1. The four analyzed soil profiles are in Tohoku area with epicentral distance up to 400 km and have increasing shear wave velocity with depth. Soil profiles have different properties: depth, number and thickness of layers, soil type and compressional to shear wave velocity in the soil. Stratigraphies and soil properties used in this analysis are shown in Tables 2-5. Soil properties are assumed uniform in each layer.

The dynamic mechanical properties of the Tohoku alluvial deposits are not provided. The normalized shear modulus decay curves employed in this work are obtained according to the hyperbolic model, as in equation (12). The applied reference strain corresponds, for each soil type in the analyzed profiles, to the 50 % reduction of shear modulus in well-known shear modulus decay curves of the literature (Tables 2-5). The curve proposed by Seed and Idriss (1970a) is used to define $\gamma_r$ for sands and the curve of Seed and Sun (1989) is applied for clays. A plasticity index in the range of PI = 20 - 40 is assumed in the relationship of Sun *et al*. (1988) to define $\gamma_r$ for volcanic ash clay and PI = 5 - 10 is adopted for silt. The reference shear strain for gravel is defined according to Seed *et al*. (1986). An almost linear behavior is assumed for stiff layers above the bedrock ($\gamma_r$ = 100 ‰). The choices of $\gamma_r$ could influence the analysis, but the variation in the dynamic response of soil columns is neglected here.

The density of soil layers in the profile NIGH11 is not provided by the KiK-Net database, so it is assumed (Table 5).

According to the proposed model, the bedrock has an elastic behavior with a high elastic modulus. The physical properties assumed for bedrock are the density $\rho_b = 2100\,\text{kg/m}^3$, the shear velocity in the bedrock $v_{sb} = 1000\,\text{m/s}$ and the pressure wave velocity $v_{pb}$ is deduced by (11), by imposing a Poisson's ratio of 0.4. The lack of geotechnical data for deeper layers induces to assume the bedrock right below the soil profile described by K-Net data.

## Input seismic signals

The four soil profiles have been selected because the vertical to horizontal peak ground acceleration ratio is higher than 70 % (Table 6), with a low compressional to shear wave velocity ratio in the soil that implies a low Poisson's ratio, according to equation (11). The minimum $v_p/v_s$ in each studied stratigraphy is indicated in Table 1. The PGA recorded at the surface of analyzed soil profiles is slightly higher than the acceleration level commonly used for structural design in high risk seismic zones. The three components of motion are recorded in North-South (NS), East-West (EW) and Up-Down (UP) directions, respectively referred to as $x$, $y$ and $z$ in the proposed model. Recorded signals have different polarization. The peak ground acceleration (PGA) and peak ground velocity (PGV) can be referred (see Table 6) to different directions of polarization (NS ≡ x or EW ≡ y). PGA and PGV are indicated by bold characters in Table 6. The three maximum acceleration components, in each direction of motion, correspond to different times. Maximum acceleration and velocity moduli at the surface of analyzed soil profiles are listed in Table 6. The waveforms are provided by the Kyoshin Network strong ground motion database (see Data and Resources Section).

Rock type profiles are selected as the sites closest to analyzed soil profiles, where accelerometer stations are placed and whose stratigraphy is defined as rock, by the K-Net database, all along the depth, until the surface ground. Rock type profiles have different epicentral distance, depth and average shear velocity in the soil, as listed in Table 7. The position of soil and rock type profiles in Tohoku area is shown in Figure 2. A thin surficial soil layer, present in some rock type profiles (Table 7), has been neglected and assimilated to rock. The lack of geotechnical data could induce to questionable results when geological homogeneity of selected rock type profiles and the underlying bedrock under analyzed soil profiles is not assessed.

404  Three-component seismic signals recorded in directions North-South, East-West and Up-Down
405  during the 9 Mw 11 March 2011 Tohoku earthquake (Table 8), at outcropping bedrock, are
406  halved and propagated in the examined soil columns FKS011, IBR007 and MYG010.
407  Acceleration signals are halved to take into account the free surface effect and integrated, to
408  obtain the corresponding input data in terms of vertically incident velocities, before being forced
409  at the base of the horizontal multilayer soil model, by the equation (3). The three components
410  induce shear loading in horizontal directions $x$ (NS) and $y$ (EW) and pressure loading in $z$-
411  direction (UD). Each signal recorded at rock sites has different amplitude and polarization. PGA
412  and PGV can be referred to different directions of polarization (PGA and PGV are indicated by
413  bold characters in Table 8).
414  Bedrock seismic records for NIGH11 (Table 8), provided by KiK-Net database (see Data and
415  Resources Section), are measured at 205 m of depth. These borehole records, assumed as
416  incident waves, are not halved before being forced at the base of the multilayer soil column.
417
418  **Validation and discussion**
419  The validation of proposed model and numerical procedure is done by comparison of computed
420  results with records in terms of surface time histories. Bedrock and surface time histories are
421  compared to investigate amplification effects in alluvial deposits.
422  A preliminary study is done for soil profiles FKS011, IBR007 and MYG010, to identify the
423  reference outcropping bedrock. In fact, a great variability of the computed surface response with
424  the choice of the rock type profile, where the input signal is recorded, is noticed, especially in
425  terms of amplitude. In Figures 3 and 4a, the various time histories of ground acceleration
426  modulus at the surface are shown for the chosen rock type profiles associated to soil profile

427  FKS011. The rock type profile where the 3C seismic record, used as incident wave, provides the

428  best numerical approximation of 3C surface record for the analyzed soil profile is identified as

429  reference outcropping bedrock for that profile.

430  Acceleration moduli are compared in Figures 3(a, c, e) and 4a for soil profile FKS011, in

431  Figures 5(a, c) and 6a for IBR007 and in Figure 7(a, c) for MYG010. The case referred as A/B is

432  associated to soil profile A with incident signal deduced halving records in rock type profile B.

433  The three acceleration components for the case of input signal recorded at the reference

434  outcropping bedrock are shown, for soil profiles FKS011, IBR007 and MYG010, in Figure 8(a,

435  b, c), respectively. Numerical results are consistent with recordings.

436  Obtained maximum accelerations are listed in Table 9 and their values are close to recorded

437  acceleration peaks (Table 6). Bold values in Table 9 correspond to selected rock type profiles

438  (reference outcropping bedrock), providing the best approximation of the acceleration modulus

439  at the surface. Bold values in Table 10 are the computed maximum velocities best reproducing

440  records. In soil profiles IBR007 and MYG010, the peak ground motion, both in terms of

441  acceleration (Table 9) and velocity (Table 10), is better reproduced by input signals recorded in

442  rock type profiles FKS031 and MYG011, respectively. The three-component signal recorded in

443  rock type profile FKS015 allows a good approximation of the maximum components and

444  modulus of acceleration in soil profile FKS011 (Table 9), while it is the signal recorded in rock

445  type profile FKS031 that better reproduces the maximum components and modulus of velocity

446  (Table 10).

447  The acceleration time history at the surface (Fig. 3(a, b)), produced by propagating the halved

448  acceleration recorded in the rock type profile FKS004 along the soil column FKS011, is not a

449  good approximation of the recorded signal. The too low average shear velocity of the rock type

450  profile FKS004, equal to 240 m/s (Table 7), could justify this inconsistency. It is important to
451  notice the variability of seismic response at the surface of a soil column with characteristics of
452  the selected rock type profile, identifying the outcropping bedrock considered in the theoretical
453  model. The shear velocity profile with depth of assumed reference rock type columns and the
454  distance between rock and soil profiles are parameters that could strongly influence the
455  numerical seismic response in soil profiles. The bedrock to surface signal amplification is shown
456  in Figures 3(b, d, f), 5(b, d) and 7(b, d) for soil profiles FKS011, IBR007 and MYG010,
457  respectively. In soil profile MYG010, the acceleration signal amplification is no so significant
458  compared with the reference bedrock signal (Fig. 7d), conversely to the other presented cases
459  (Figs 4b and 7b).
460  Seismic response at the surface of soil profile NIGH11 is shown in Figure 9 in terms of
461  maximum acceleration modulus. Numerical acceleration is slightly amplified compared with
462  records. Further investigations could be undertaken by imposing a borehole boundary condition
463  (instead of absorbing boundary condition (3)), at the soil-bedrock interface of the numerical
464  model, to observe if this effect persist.
465  The assumption of soil density in NIGH11, not provided by KiK-Net database, could also
466  influence the seismic site response.
467
468  **1D-3C VS 1D-1C APPROACH**
469  The seismic response of a horizontally multilayered soil to the propagation of a three-component
470  signal (1D-3C approach) is compared in the case of the 2011 Tohoku earthquake, to the
471  superposition of the three independently propagated components (1D-1C approach). The shear
472  modulus decreases, in the case of 1C propagation, according to the shear modulus decay curve

473  of the material obtained by laboratory tests. The stress-strain curve during a loading is referred
474  to a backbone curve, obtained knowing the shear modulus decay curve.
475  Modeling the one-directional propagation of a three-component earthquake allows to take into
476  account the interactions between shear and pressure components of the seismic load. Nonlinear
477  and multiaxial coupling effects appear under a triaxial stress state induced by a cyclic 3D
478  loading.
479  The comparison between 1D-1C and 1D-3C approaches is shown in Figure 10 for soil profiles
480  FKS011 and IBR007, respectively, in terms of surface time histories. Stratigraphies and soil
481  properties are given in Tables 2 and 3. The interaction between multiaxial stresses in the 3C
482  approach yields a reduction of the ground motion at the surface. The modulus of acceleration at
483  the outcropping bedrock appears amplified at the surface of analyzed soil columns for both 1D-
484  1C and 1D-3C approaches, but peak accelerations are reduced in 1D-3C case and closer to
485  records (Table 9). The PGV appears also reduced in the 1D-3C case, compared with the 1D-1C
486  approach (Table 10).
487  The local response to a three-component earthquake in soil profiles FKS011 and IBR007 is
488  analyzed in terms of depth profiles of maximum acceleration and velocity modulus and
489  maximum octahedral stress and strain and in terms of stress-strain cycles in the most deformed
490  layer (Figs 11 and 12).
491  The maximum motion modulus profile with depth shows, at each $z$-coordinate, the maximum
492  modulus of the ground motion during shaking. The maximum acceleration modulus profiles with
493  depth are displayed in these figures without low-pass filtering operations. Equation (13) is used
494  to evaluate octahedral strains and stresses, which maximum values during the loading time are
495  represented as profiles with depth. Hysteresis loops, at a given depth, are shown in terms of shear strain and stress.

Maximum accelerations and velocities appear slightly higher for the combination of three 1C-propagations (1D-1C approach). Maximum stresses are reduced, in the 1D-3C case, and in softer layers maximum strains can be higher.

Cyclic shear strains with amplitude higher than the elastic behavior range limit give open loops in the shear stress-shear strain plane, exhibiting strong hysteresis. Due to nonlinear effects, the shear modulus decreases and the dissipation increases with increasing strain amplitude. The soil column cyclic responses in terms of shear stress and strain in $x$-direction when it is affected by a triaxial input signal (1D-3C) and when the $x$-component of the input signal is independently propagated (1D-1C) are compared in Figures 11(b, c) and 12(b, c). From one to three components, for a given maximum strain amplitude, the shear modulus decreases and the dissipation increases. Under triaxial loading the material strength is lower than for simple shear loading, referred to as the backbone curve.

Hysteresis loops for each horizontal direction are altered as a consequence of the interaction between loading components. This result confirms the findings of the parametric analysis using synthetic wavelets by Santisi d'Avila *et al.* (2012). In the case of one-component loading, the shape of the first loading curve is the same as the backbone curve and the shape of hysteresis loops remains unvaried at each cycle, for shear strains in the range of stable nonlinearity. In the case of three-component loading, the shape of the hysteresis loops changes at each cycle, also in a strain range that in the case of 1C loading is of stable nonlinearity, because the shape of loops is disturbed by the multiaxial stress coupling.

The main difference between 1D-1C and 1D-3C approach is remarkable in terms of ground motion time history, maximum stress and hysteretic behavior, with more nonlinearity and coupling effects between components.

**1D-3C LOCAL SEISMIC RESPONSE ANALYSIS IN THE TOHOKU AREA**

This research aims to provide a tool to study the local seismic response in case of strong earthquakes affecting alluvial sites. The proposed model allows to preview possible amplifications of seismic motion at the surface, influenced by stratigraphic characteristics, and to evaluate non-measured parameters of motion, stress and strain along the soil profiles, in order to investigate nonlinear effects in deeper detail. Depth profiles of maximum acceleration and velocity modulus, maximum octahedral stress and strain are shown in Figures 11a, 12a and 13a, for soil profiles FKS011, IBR007, MYG010, respectively. The results for soil profile NIGH11 are shown in Figure 14.

Soft layers and high strain drops at layer interfaces can be identified evaluating the maximum strain profiles with depth. We observe that maximum strains along the soil profile are present in layer interfaces (Figs 11a, 12a, 13a and 14).

The 1D-3C approach allows to evaluate non-measured parameters of motion, stress and strain along the analyzed soil profile, influenced by the input motion polarization and 3D loading path. Non null strain and stress components are assessed along the soil profile, namely the three strains in $z$-direction, $\gamma_{yz}$, $\gamma_{yz}$ and $\varepsilon_{zz}$, and consequent stresses $\sigma_{xx}$, $\sigma_{yy}$, $\tau_{yz}$, $\tau_{zx}$ and $\sigma_{zz}$.

The shape of the shear stress-strain cycles in $x$-direction (respectively $y$-direction) reflects coupling effects with loads in directions $y$ (respectively $x$) and $z$. At a given depth, nonlinear effects are more important for the minimum peak horizontal component that is the most influenced by three-dimensional motion coupling (Figs 11c, 12c and 13b).

In particular for the Tohoku earthquake, we detect, in all hysteresis loops (Figs 11(b, c), 12(b, c)

542  and 13(b, c)), two successive events (Bonilla *et al*., 2011). This earthquake feature is also
543  observed in a time-frequency polarization analysis. Stockwell amplitude spectra of separate
544  horizontal acceleration components at the surface are compared in Figure 15, for records (up)
545  and numerical computations (down) in *x*- (Fig. 15a) and *y*-direction (Fig. 15b). Two successive
546  events can be easily distinguished, the range of frequencies involved throughout the time is
547  coherent and spectral amplitudes are similar for given time and frequency. That confirms the
548  reliability of the proposed model. It will be interesting to investigate, in a future study, the
549  different response of a soil column to two independent and successive events.
550  In Figure 13b, we can remark a completely negligible overtaking of the one-dimensional soil
551  strength (backbone curve). This numerical error of the three-dimensional soil behavior routine,
552  due to convergence difficulties, becomes more evident for strains higher than about 5 %, when
553  the constitutive model gets to be unusable (Lenti, 2006). The implemented MPII type model
554  gives reliable results in a range of stable nonlinearity. Liquefaction problems cannot be
555  investigated. Being the proposed propagation model totally independent of the applied
556  constitutive relation, a major goal is to implement a relation for saturated soils.
557  The variability of seismic response at the surface of soil columns with the characteristics of
558  selected rock type profiles, approximating the outcropping bedrock, demands future statistical
559  studies to analyze the local seismic response of a site accounting for various rock profiles and
560  different earthquake records.
561
562  **CONCLUSIONS**
563  A one-dimensional three-component geomechanical model is proposed and discussed, to analyze
564  the propagation of 3C seismic waves due to the strong quakes in 1D soil profiles (1D-3C

approach). A promising solution for strong seismic motion evaluation and site effect analysis is provided.

A three-dimensional constitutive relation of the Masing-Prandtl-Ishlinskii-Iwan (MPII) type, for cyclic loading, is implemented in a finite element scheme, modeling a horizontally layered soil. The adopted rheological model for soils has been selected for its 3D features with nonlinear behavior for both loading and unloading and, above all, because few parameters are necessary to characterize the soil hysteretic behavior.

The analysis of four soil profiles in the Tohoku area (Japan), shaken by the 9 Mw 11 March 2011 Tohoku earthquake, is presented in this paper. The validation of the 1D-3C approach against recorded surface time histories is carried out and the reliability of the proposed model is confirmed.

We selected, in this study, some rock type profiles close to analyzed soil profiles and we use as incident loading the halved signal recorded at rock outcrops. The variability of the surface ground motion with the bedrock incident loading is observed. The signal recorded in outcropping bedrock, permitting to obtain the best approximation of the surface seismic record is assumed as reference bedrock motion for the analyzed soil profile. The lack of geotechnical data could induce to questionable results when geological homogeneity of selected rock type outcrops and the modeled bedrock underlying analyzed multilayered soils is not assessed. More quantitative analyses could be undertaken when more available input data will permit to increase the accuracy of results. Statistical studies using records of different earthquakes at a same site could be undertaken using the 1D-3C approach for the evaluation of local seismic response for site effect analyses.

The combination of three separate 1D-1C nonlinear analyses is compared to the proposed 1D-3C

approach. Motion amplification effects at the surface are reduced in the 1D-3C approach due to nonlinearities and three-dimensional motion coupling. Multiaxial stress states induce strength reduction of the material and larger damping effects. The shape of hysteresis loops changes at each cycle in the 1D-3C approach, also in a strain range that in the case of one-component loading is of stable nonlinearity.

Effects of the input motion polarization and 3D loading path can be detected by the 1D-3C approach, that allow to evaluate non-measured parameters of motion, stress and strain along the analyzed soil profile, in order to detail nonlinear effects. Soil properties such as the Poisson's ratio have great impact on local seismic response, influencing the soil dissipative properties. Input motion properties such as the polarization (vertical to horizontal component ratio) affect energy dissipation rate and the amplification effect. In particular, a low seismic velocity ratio in the soil and a high vertical to horizontal component ratio increase the three-dimensional mechanical interaction and progressively change the hysteresis loop size and shape at each cycle. Maximum strains are induced in layer interfaces, where waves encounter large variations of impedance contrast, along the soil profile. Nonlinearity effects are more important in the direction of minimum peak horizontal component that is the most influenced by three-dimensional motion coupling.

In particular for the 2011 Tohoku earthquake, the two successive events, detected by records, are numerically reproduced (hysteresis loops, Stockwell amplitude spectra).

The extension of the proposed 1D-3C approach to higher strain rates is planned as further investigation to be able to study the effects of soil nonlinearity in saturated conditions.

## DATA AND RESOURCES

Seismograms and soil stratigraphic setting used in this study are provided by the National research Institute for Earth science and Disaster prevention (NIED), in Japan, and can be obtained from the Kyoshin and Kiban-Kyoshin Networks at www.k-net.bosai.go.jp (last accessed May 2012).

## ACKNOWLEDGMENTS

We are grateful to Florent De Martin as well as an anonymous reviewer, for their careful revision of this manuscript and their constructive suggestions.

721  **AUTHORS' AFFILIATION**

722  Maria Paola Santisi d'Avila

723  Laboratoire Jean Alexandre Dieudonné

724  University of Nice Sophia Antipolis

725  Parc Valrose

726  Nice, France 06108

727

728  Jean-François Semblat and Luca Lenti

729  IFSTTAR

730  University Paris-Est

731  14 Boulevard Newton

732  Marne la Vallée, France 77447

733

734

735

736

737

738

739

740

741

742

743

744  **TABLES**

746  **Table 1.** Selected soil profiles in Tohoku area (Japan)

| Site name - Prefecture | Site code | Epicentral distance (km) | Depth H (m) | Average $v_s$ (m/s) | min $\{v_p/v_s\}$ |
|---|---|---|---|---|---|
| IWAKY - FUKUSHIMAKEN | FKS011 | 206 | 10.00 | 222 | 3.05 |
| NAKAMINATO - IBARAKIKEN | IBR007 | 279 | 20.35 | 239 | 2.30 |
| ISHINOMAKI - MIYAGIKEN | MYG010 | 143 | 20.45 | 247 | 4.62 |
| KAWANISHI - NIIGATAKEN | NIGH11 | 378 | 205.0 | 578 | 2.45 |

749  **Table 2.** Stratigraphy and soil properties of profile FKS011

| FKS011 | H-z (m) | th (m) | $\rho$ (kg/m$^3$) | $v_s$ (m/s) | $v_p$ (m/s) | $\gamma_r$ (‰) |
|---|---|---|---|---|---|---|
| Fill soil | 2.2 | 2.2 | 1430 | 100 | 700 | 0.800 |
| Silt | 3 | 0.8 | 1650 | 210 | 700 | 0.427 |
| | 4 | 1 | 1720 | 210 | 1300 | 0.427 |
| | 5.95 | 1.95 | 1660 | 330 | 1300 | 0.427 |
| Clay | 6.85 | 0.9 | 1810 | 330 | 1300 | 2.431 |
| Rock | 8 | 1.15 | 1970 | 330 | 1300 | 100 |
| | 9 | 1 | 1980 | 590 | 1800 | 100 |
| | 10 | 1 | 2060 | 590 | 1800 | 100 |

752 **Table 3.** Stratigraphy and soil properties of profile IBR007

| IBR007 | H-z (m) | th (m) | $\rho$ (kg/m$^3$) | $v_s$ (m/s) | $v_P$ (m/s) | $\gamma_r$ (‰) |
|---|---|---|---|---|---|---|
| Fill soil | 2 | 2 | 1450 | 80 | 260 | 1.065 |
|  | 3.9 | 1.9 | 1750 | 150 | 520 | 1.065 |
| Volcanic ash clay | 4.4 | 0.5 | 1810 | 150 | 520 | 1.065 |
| Sand | 6 | 1.6 | 1910 | 200 | 1220 | 0.368 |
|  | 7.8 | 1.8 | 1850 | 200 | 1220 | 0.368 |
|  | 9 | 1.2 | 1770 | 200 | 1220 | 0.427 |
| Silt | 10 | 1 | 1810 | 530 | 1220 | 0.427 |
|  | 11.2 | 1.2 | 1920 | 530 | 1220 | 0.427 |
| Sand | 12.7 | 1.5 | 1980 | 530 | 1220 | 0.368 |
| Gravel | 14.1 | 1.4 | 2060 | 530 | 1220 | 0.143 |
| Clay | 15.1 | 1 | 1880 | 530 | 1220 | 2.431 |
|  | 16 | 0.9 | 1960 | 610 | 1920 | 0.368 |
| Sand | 17 | 1 | 1880 | 610 | 1920 | 0.368 |
|  | 20.35 | 3.35 | 1900 | 610 | 1920 | 0.368 |

753

754

755 **Table 4.** Stratigraphy and soil properties of profile MYG010

| MYG010 | H-z (m) | th (m) | $\rho$ (kg/m$^3$) | $v_s$ (m/s) | $v_P$ (m/s) | $\gamma_r$ (‰) |
|---|---|---|---|---|---|---|
| Fill soil | 1.5 | 1.5 | 1600 | 100 | 280 | 0.368 |
|  | 2 | 0.5 | 1660 | 150 | 1480 | 0.368 |
|  | 3 | 1 | 1810 | 150 | 1480 | 0.368 |
|  | 4 | 1 | 1950 | 150 | 1480 | 0.368 |
|  | 5 | 1 | 1900 | 320 | 1480 | 0.368 |
| Sand | 6 | 1 | 1860 | 320 | 1480 | 0.368 |
|  | 7 | 1 | 1900 | 320 | 1480 | 0.368 |
|  | 8 | 1 | 1810 | 320 | 1480 | 0.368 |
|  | 17 | 9 | 1890 | 300 | 1480 | 0.368 |
|  | 20.45 | 3.45 | 1850 | 300 | 1480 | 0.368 |

756

757

Table 5. Stratigraphy and soil properties of profile NIGH11

| NIGH11 | H-z (m) | th (m) | $\rho$ (kg/m$^3$) | $v_s$ (m/s) | $v_p$ (m/s) | $\gamma_r$ (‰) |
|---|---|---|---|---|---|---|
| Fill soil | 2 | 2 | 1800 | 200 | 500 | 0.143 |
| Gravel | 30 | 28 | 1800 | 400 | 1830 | 0.143 |
| Rock | 46 | 16 | 1900 | 400 | 1830 | 100 |
| Silt | 57 | 11 | 1900 | 400 | 1830 | 0.427 |
| | 63 | 6 | 1900 | 700 | 1830 | 100 |
| Rock | 85 | 22 | 1900 | 520 | 1830 | 100 |
| | 185 | 100 | 1900 | 650 | 1830 | 100 |
| Gravel | 198 | 13 | 1800 | 850 | 2080 | 0.143 |
| Rock | 205 | 7 | 2000 | 850 | 2080 | 100 |

Table 6. Acceleration and velocity recorded at the surface of selected soil profiles during the 2011 Tohoku earthquake

| Site code | $a_x$ (m/s$^2$) | $a_y$ (m/s$^2$) | $a_z$ (m/s$^2$) | $|a|$ (m/s$^2$) | $a_z$/max{$a_x$, $a_y$} (%) | $v_x$ (m/s) | $v_y$ (m/s) | $v_z$ (m/s) | $|v|$ (m/s) | $v_z$/max{$v_x$, $v_y$} (%) |
|---|---|---|---|---|---|---|---|---|---|---|
| FKS011 | **3.74** | 3.12 | 3.00 | **4.47** | 80 | **0.39** | 0.34 | 0.12 | **0.47** | 31 |
| IBR007 | **5.43** | 5.10 | 4.12 | **5.87** | 76 | 0.29 | **0.44** | 0.13 | **0.49** | 30 |
| MYG010 | **4.58** | 3.77 | 3.32 | **4.88** | 72 | 0.50 | **0.56** | 0.16 | **0.68** | 29 |
| NIGH11 | **0.22** | 0.18 | 0.16 | **0.26** | 73 | 0.050 | **0.056** | 0.041 | **0.058** | 73 |

Table 7. Selected rock type profiles in Tohoku area (Japan)

| Site name | Prefecture | Site code | Epicentral distance (km) | Depth H (m) | Average $v_s$ (m/s) | Surface soil depth (m) |
|---|---|---|---|---|---|---|
| IITATE | FUKUSHIMAKEN | FKS004 | 193 | 10.42 | 240 | 0.50 |
| TANAGURA | FUKUSHIMAKEN | FKS015 | 250 | 10.03 | 463 | 0.50 |
| NIHOMMATSU | FUKUSHIMAKEN | FKS019 | 220 | 11.27 | 1025 | 0.20 |
| KAWAUCHI | FUKUSHIMAKEN | FKS031 | 199 | 10.11 | 437 | - |
| OHFUNATO | IWATEKEN | IWT008 | 148 | 10.00 | 750 | 0.15 |
| OSHIKA | MIYAGIKEN | MYG011 | 121 | 20.00 | 1220 | 0.05 |
| UTSUNOMIYA | TOCHIGIKEN | TCG007 | 314 | 10.14 | 388 | 2.30 |

Table 8. Acceleration and velocity recorded at the surface of selected rock type profiles and borehole acceleration and velocity recorded in soil profile NIGH11, during the 2011 Tohoku earthquake

| Site code | $a_x$ (m/s$^2$) | $a_y$ (m/s$^2$) | $a_z$ (m/s$^2$) | $|a|$ (m/s$^2$) | $a_z/\max\{a_x, a_y\}$ (%) | $v_x$ (m/s) | $v_y$ (m/s) | $v_z$ (m/s) | $|v|$ (m/s) | $v_z/\max\{v_x, v_y\}$ (%) |
|---|---|---|---|---|---|---|---|---|---|---|
| FKS004 | **2.98** | 2.53 | 1.49 | 3.53 | 50 | **0.21** | 0.17 | 0.08 | 0.23 | 38 |
| FKS015 | **1.36** | 1.01 | 0.58 | 1.42 | 43 | **0.17** | 0.16 | 0.10 | 0.18 | 59 |
| FKS019 | 2.07 | **2.16** | 0.84 | 2.29 | 39 | 0.27 | **0.30** | 0.13 | 0.30 | 44 |
| FKS031 | **2.34** | 2.17 | 1.43 | 2.40 | 61 | **0.34** | 0.29 | 0.12 | 0.37 | 35 |
| IWT008 | 1.26 | **1.66** | 0.61 | 2.03 | 37 | 0.10 | **0.14** | 0.09 | 0.17 | 64 |
| MYG011 | **4.39** | 3.26 | 1.24 | 4.42 | 28 | 0.19 | **0.37** | 0.16 | 0.38 | 43 |
| TCG007 | 0.81 | **0.86** | 0.60 | 0.98 | 70 | 0.19 | **0.14** | 0.09 | 0.19 | 47 |
| NIGH11 | **0.14** | 0.14 | 0.13 | 0.15 | 96 | 0.042 | **0.058** | 0.039 | 0.059 | 67 |

Table 9. Numerical acceleration evaluated at the surface of selected soil profiles

| Soil profile site code | Rock profile site code | $a_x$ (m/s²) | $a_y$ (m/s²) | $a_z$ (m/s²) | $|a|$ (m/s²) 1D-3C | 1D-1C |
|---|---|---|---|---|---|---|
| FKS011 | FKS004 | 5.99 | 5.50 | 2.94 | 5.68 | |
| FKS011 | **FKS015** | 3.78 | 3.92 | 1.64 | **4.55** | 5.72 |
| FKS011 | FKS019 | 4.66 | 5.06 | 1.68 | 4.76 | |
| FKS011 | FKS031 | 4.97 | 4.50 | 2.78 | 4.99 | |
| IBR007 | FKS015 | 3.73 | 3.21 | 2.21 | 3.95 | |
| IBR007 | **FKS031** | 5.59 | 5.45 | 2.73 | **6.07** | 7.54 |
| IBR007 | TCG007 | 3.04 | 3.05 | 2.09 | 3.45 | |
| MYG010 | IWT008 | 3.11 | 2.91 | 3.11 | 3.23 | |
| MYG010 | **MYG011** | 4.08 | 3.75 | 3.43 | **4.85** | |
| NIGH11 | **NIGH11** | 0.33 | 0.38 | 0.28 | **0.39** | |

Table 10. Numerical velocity evaluated at the surface of selected soil profiles

| Soil profile site code | Rock profile site code | $v_x$ (m/s) | $v_y$ (m/s) | $v_z$ (m/s) | $|v|$ (m/s) 1D-3C | 1D-1C |
|---|---|---|---|---|---|---|
| FKS011 | FKS004 | 0.32 | 0.25 | 0.08 | 0.33 | |
| | **FKS015** | 0.25 | 0.23 | 0.10 | 0.25 | 0.26 |
| | FKS019 | 0.37 | 0.42 | 0.13 | 0.43 | |
| | **FKS031** | **0.43** | **0.38** | **0.12** | **0.48** | |
| IBR007 | FKS015 | 0.21 | 0.25 | 0.11 | 0.28 | |
| | **FKS031** | **0.39** | **0.38** | **0.15** | **0.48** | 0.52 |
| | TCG007 | 0.26 | 0.18 | 0.10 | 0.26 | |
| MYG010 | IWT008 | 0.16 | 0.20 | 0.09 | 0.24 | |
| | **MYG011** | **0.17** | **0.42** | **0.16** | **0.45** | |
| NIGH11 | **NIGH11** | 0.11 | 0.15 | 0.08 | 0.15 | |

**FIGURE CAPTIONS**

**Figure 1.** Spatial discretization of a horizontally layered soil forced at its base by a halved three-component earthquake, recorded at a close outcropping bedrock site.

**Figure 2.** Geographical position of analyzed K-Net stations, placed at the surface of soil (bold) and rock type (italic) profiles, in the Tohoku area (Japan).

**Figure 3.** Time history of acceleration modulus during Tohoku earthquake: measured data and numerical solution at the ground surface (a, c, e); reference bedrock signal and surface numerical solution (b, d, f), for cases FKS011/FKS004 (a,b), FKS011/FKS019 (c,d) and FKS011/FKS031 (e, f).

**Figure 4.** Time history of acceleration modulus during Tohoku earthquake: measured data and numerical solution at the ground surface (a); reference bedrock signal and surface numerical solution (b), for case FKS011/FKS015.

**Figure 5.** Time history of acceleration modulus during Tohoku earthquake: measured data and numerical solution at the ground surface (a, c); reference bedrock signal and surface numerical solution (b, d), for cases IBR007/FKS015 (a,b) and IBR007/TCG007 (c,d).

**Figure 6.** Time history of acceleration modulus during Tohoku earthquake: measured data and numerical solution at the ground surface (a); reference bedrock signal and surface numerical solution (b), for case IBR007/FKS031.

**Figure 7.** Time history of acceleration modulus during Tohoku earthquake: measured data and numerical solution at the ground surface (a, c); reference bedrock signal and surface numerical solution (b, d), for cases MYG010/IWT008 (a,b) and MYG010/MYG011 (c,d).

**Figure 8.** Three-component acceleration time history at the ground surface during Tohoku

801 earthquake: measured data and numerical solution in directions x (left), y (middle) and z (right),

802 for cases FKS011/FKS015 (a), IBR007/FKS031 (b) and MYG010/MYG011 (c).

803 **Figure 9.** Time history of acceleration modulus during Tohoku earthquake: measured data and

804 numerical solution at the ground surface (a); reference bedrock signal and surface numerical

805 solution (b), for soil profile NIGH11.

806 **Figure 10.** Time history of acceleration modulus at the ground surface during Tohoku

807 earthquake: 1D-3C and 1D-1C numerical solutions for cases FKS011/FKS015 (a) and

808 IBR007/FKS031 (b).

809 **Figure 11.** 1D-3C and 1D-1C seismic response during the Tohoku earthquake, for the case

810 FKS011/FKS015: profiles of maximum acceleration and velocity modulus, octahedral strain

811 and stress with depth (a); shear stress-strain loops at 2 m depth in x- (b) and y-direction (c).

812 **Figure 12.** 1D-3C and 1D-1C seismic response during the Tohoku earthquake, for the case

813 IBR007/FKS031: profiles of maximum acceleration and velocity modulus, octahedral strain and

814 stress with depth (a); shear stress-strain loops at 8.5 m depth in x- (b) and y-direction (c).

815 **Figure 13.** 1D-3C and 1D-1C seismic response during the Tohoku earthquake, for the case

816 MYG010/MYG011: profiles of maximum acceleration and velocity modulus, octahedral strain

817 and stress with depth (a); shear stress-strain loops at 3.5 m depth in x- (b) and y-direction (c).

818 **Figure 14.** Maximum acceleration, velocity, octahedral strain and stress profiles with depth in

819 soil profile NIGH11 during 2011 Tohoku earthquake.

820 **Figure 15.** Spectral amplitude variation with time and frequency at the ground surface, in

821 horizontal directions x (a) and y (b), during the Tohoku earthquake, evaluated using measured

822 acceleration (up) and computed acceleration (down) as input, for the case MYG010/MYG011.

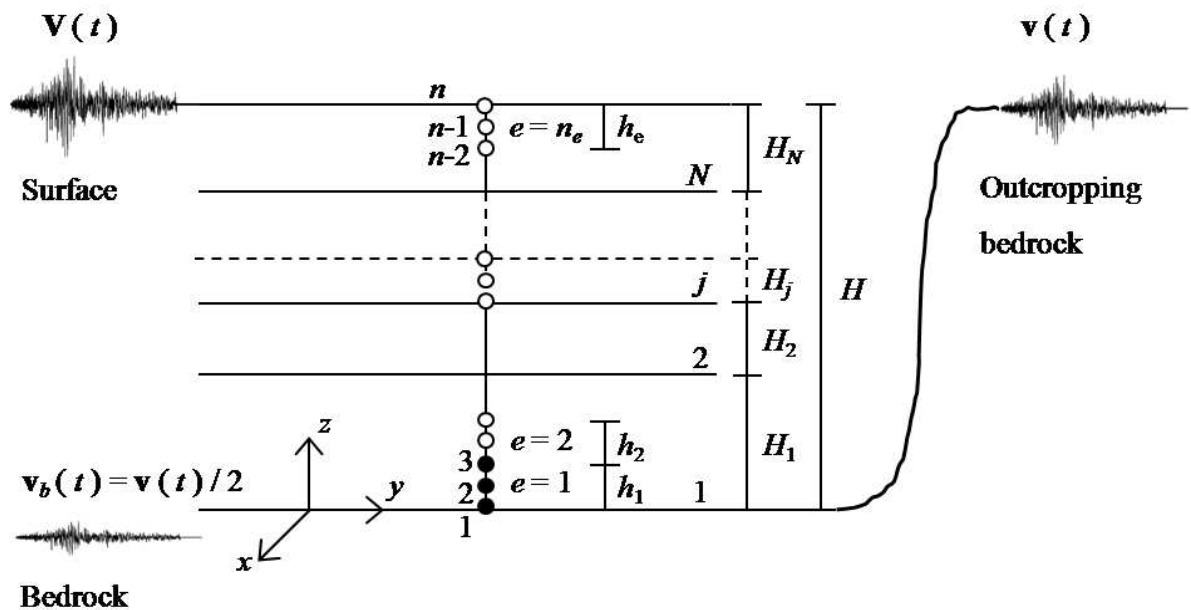

**Figure 1.** Spatial discretization of a horizontally layered soil forced at its base by a halved three-component earthquake, recorded at a close outcropping bedrock site.

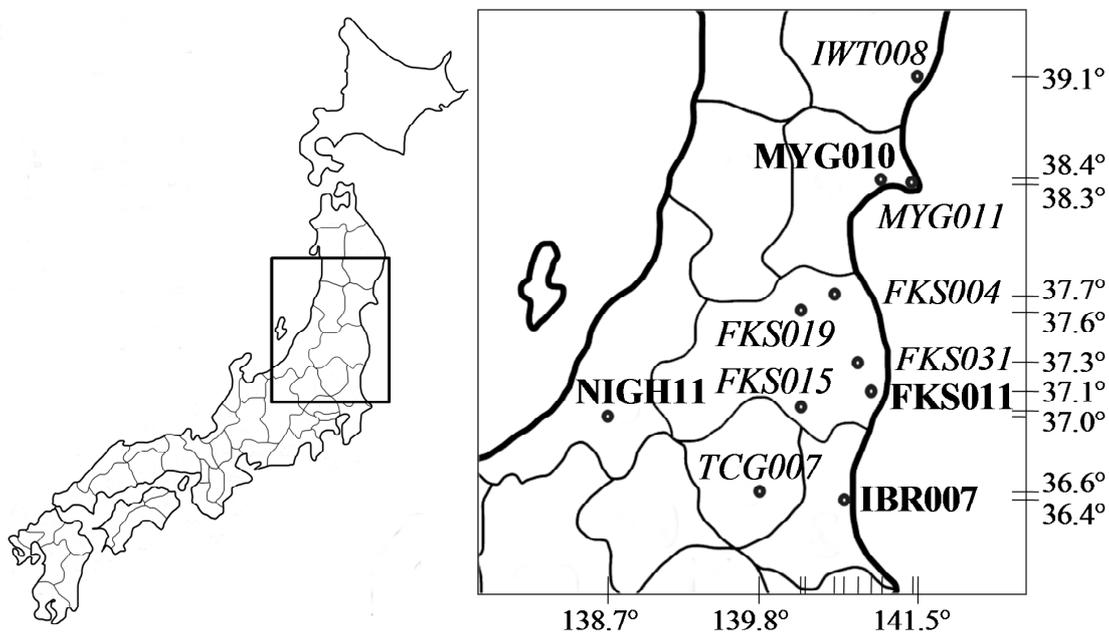

**Figure 2.** Geographical position of analyzed K-Net stations, placed at the surface of soil (bold) and rock type (italic) profiles, in the Tohoku area (Japan).

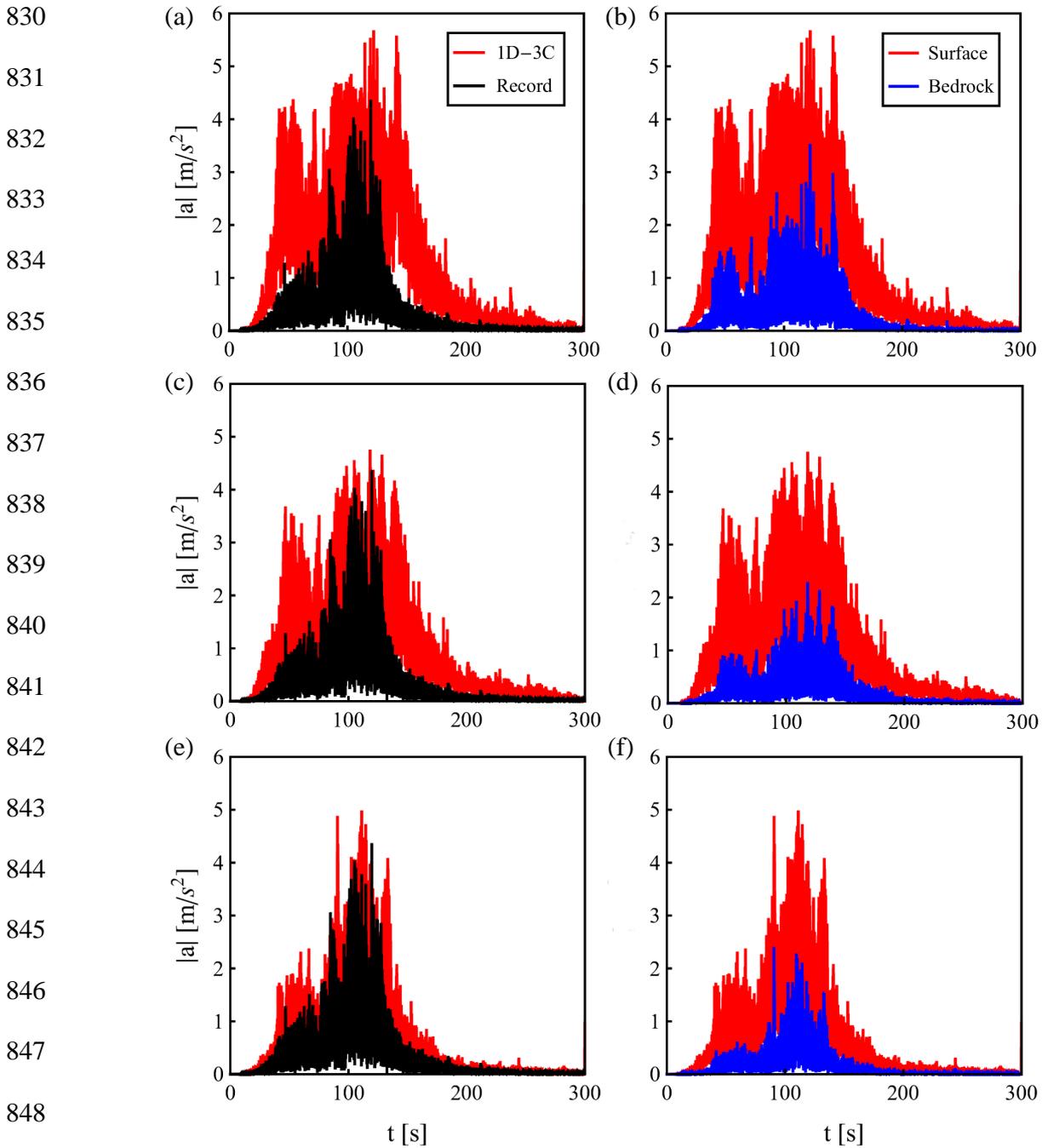

**Figure 3.** Time history of acceleration modulus during Tohoku earthquake: measured data and numerical solution at the ground surface (a, c, e); reference bedrock signal and surface numerical solution (b, d, f), for cases FKS011/FKS004 (a,b), FKS011/FKS019 (c,d) and FKS011/FKS031 (e, f).

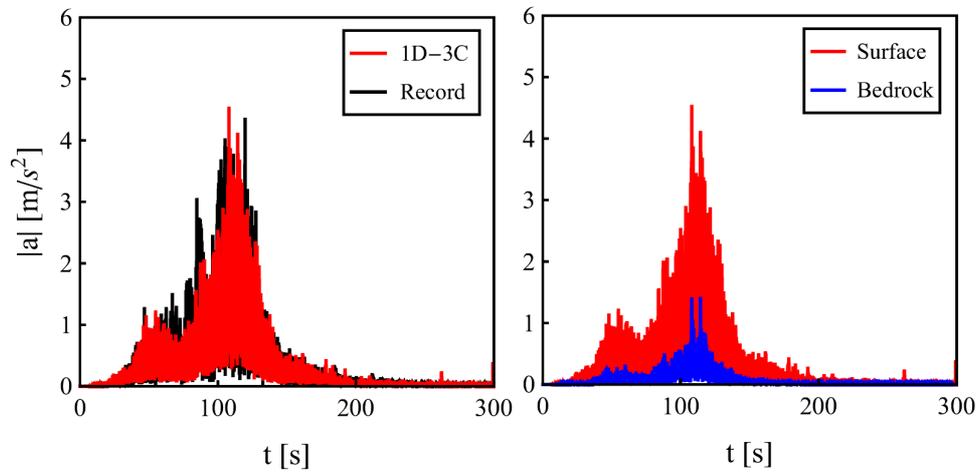

**Figure 4.** Time history of acceleration modulus during Tohoku earthquake: measured data and numerical solution at the ground surface (a); reference bedrock signal and surface numerical solution (b), for case FKS011/FKS015.

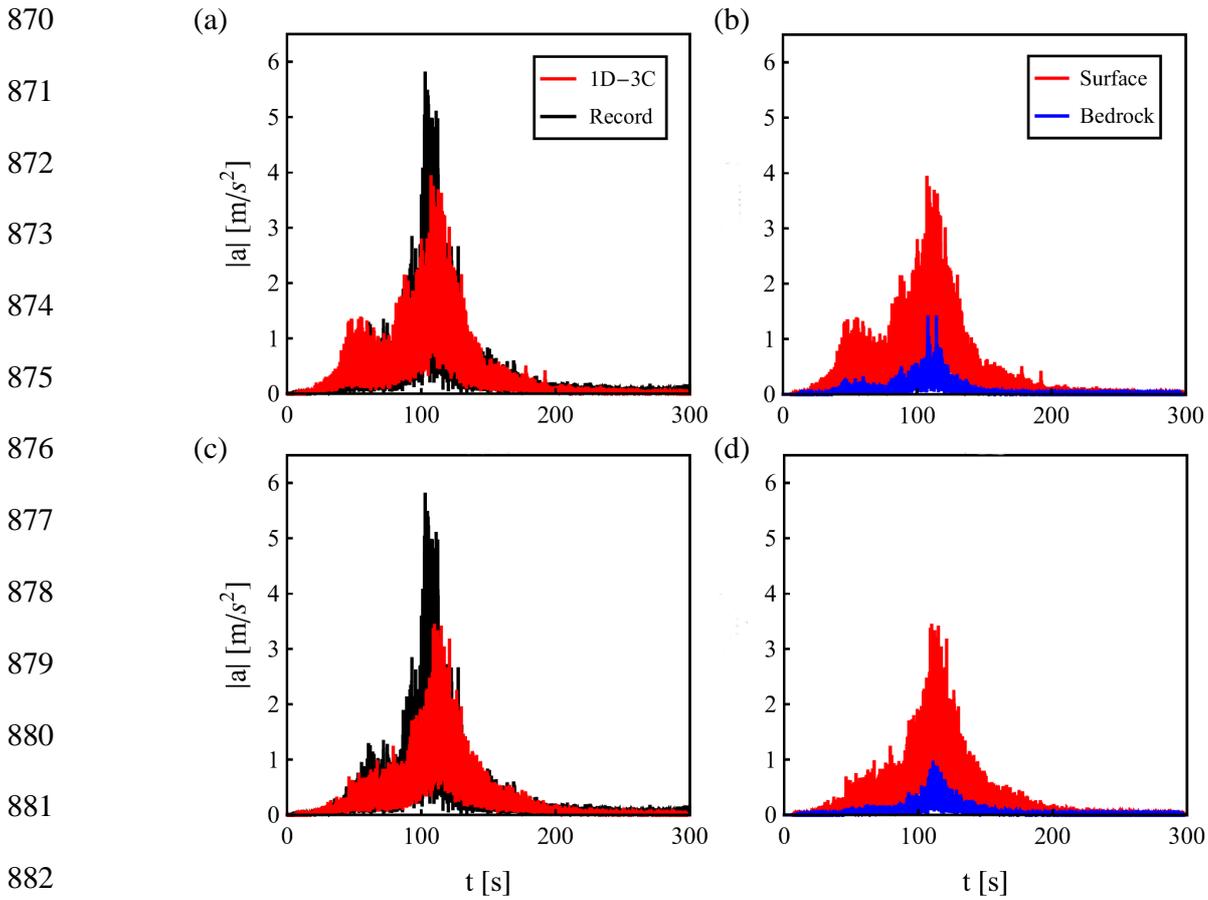

**Figure 5.** Time history of acceleration modulus during Tohoku earthquake: measured data and numerical solution at the ground surface (a, c); reference bedrock signal and surface numerical solution (b, d), for cases IBR007/FKS015 (a,b) and IBR007/TCG007 (c,d).

(a) (b)

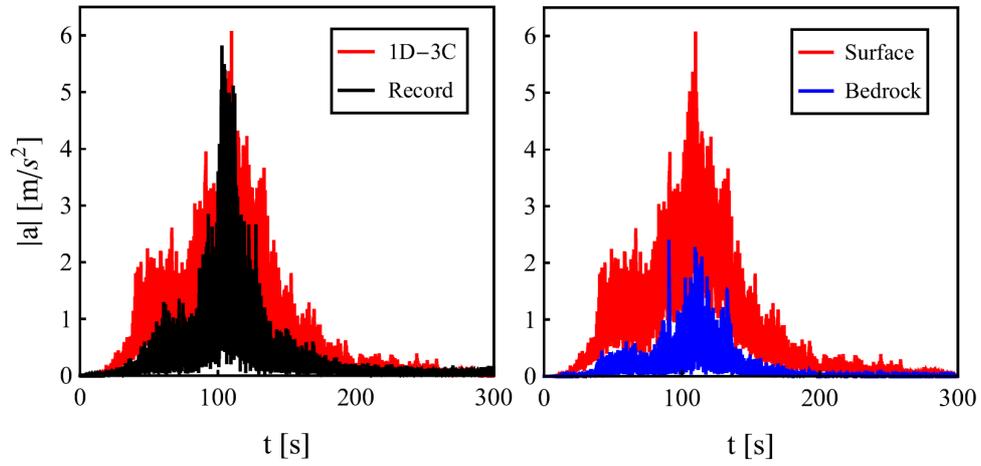

**Figure 6.** Time history of acceleration modulus during Tohoku earthquake: measured data and numerical solution at the ground surface (a); reference bedrock signal and surface numerical solution (b), for case IBR007/FKS031.

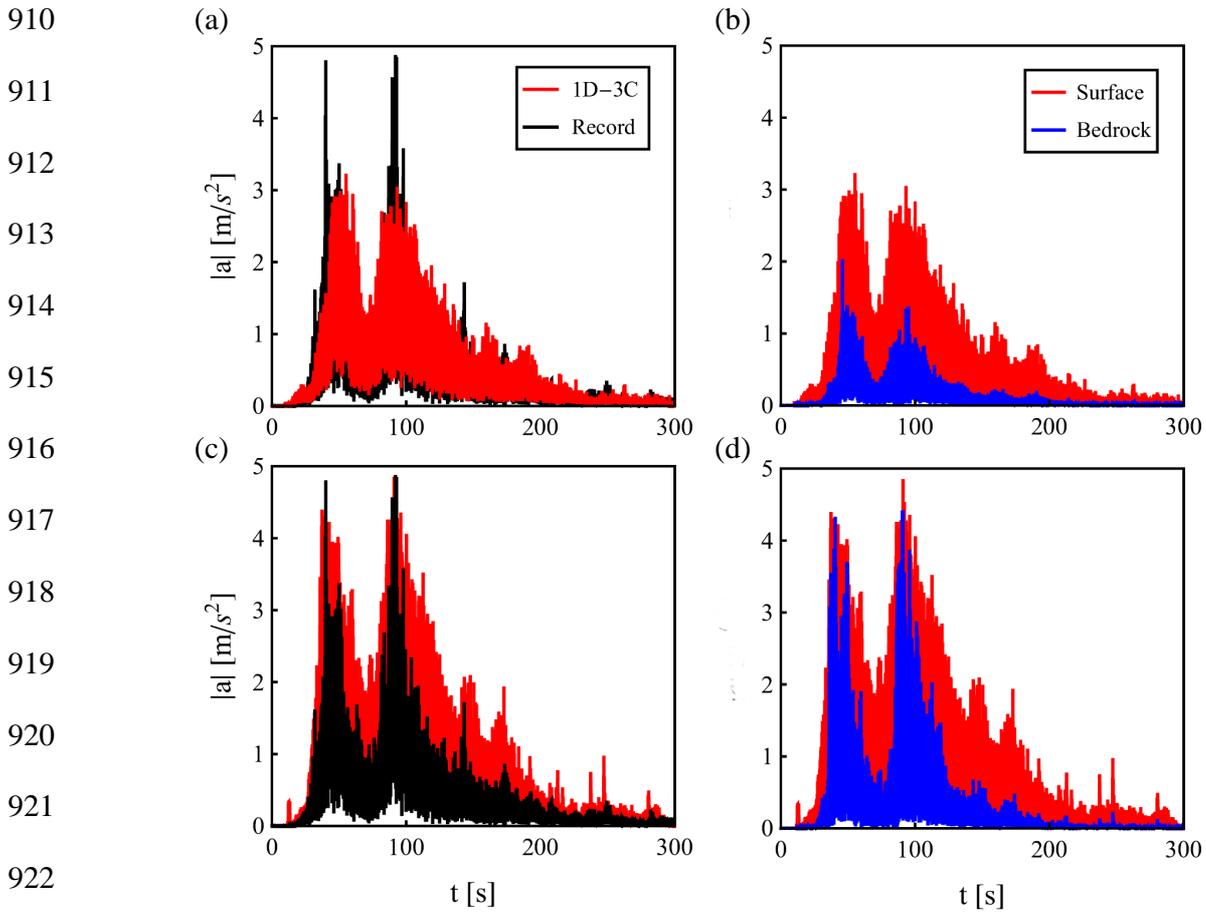

**Figure 7.** Time history of acceleration modulus during Tohoku earthquake: measured data and numerical solution at the ground surface (a, c); reference bedrock signal and surface numerical solution (b, d), for cases MYG010/IWT008 (a,b) and MYG010/MYG011 (c,d).

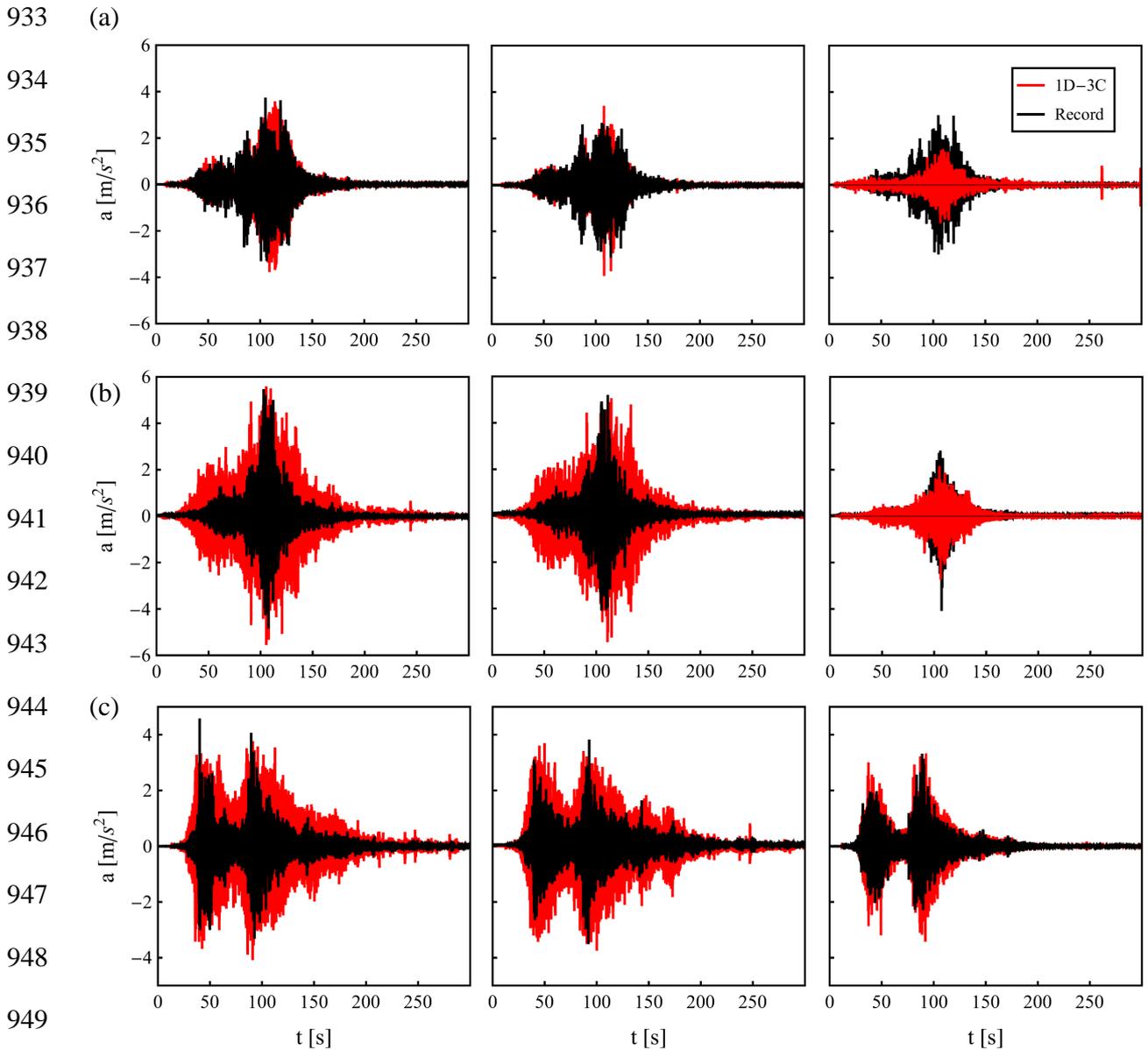

**Figure 8.** Three-component acceleration time history at the ground surface during Tohoku earthquake: measured data and numerical solution in directions x (left), y (middle) and z (right), for cases FKS011/FKS015 (a), IBR007/FKS031 (b) and MYG010/MYG011 (c).

956     (a)        (b)

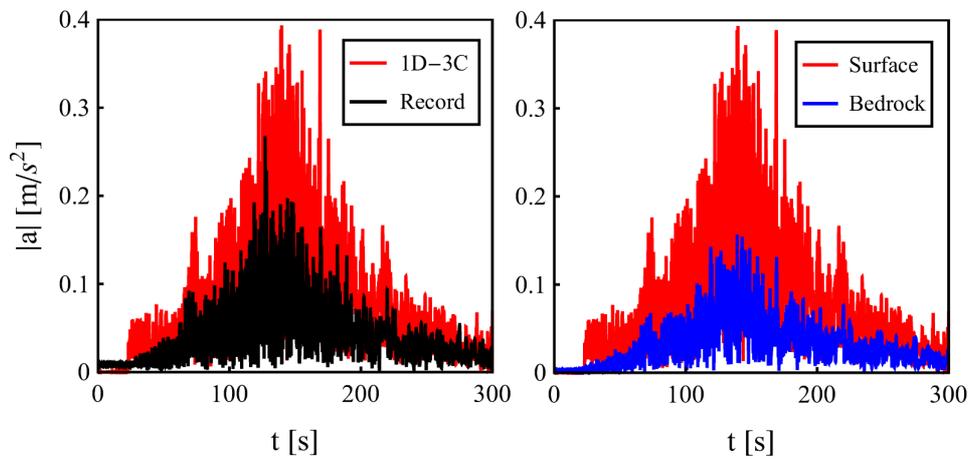

957

958 **Figure 9.** Time history of acceleration modulus during Tohoku earthquake: measured data and

959 numerical solution at the ground surface (a); reference bedrock signal and surface numerical

960 solution (b), for soil profile NIGH11.

961

962     (a)        (b)

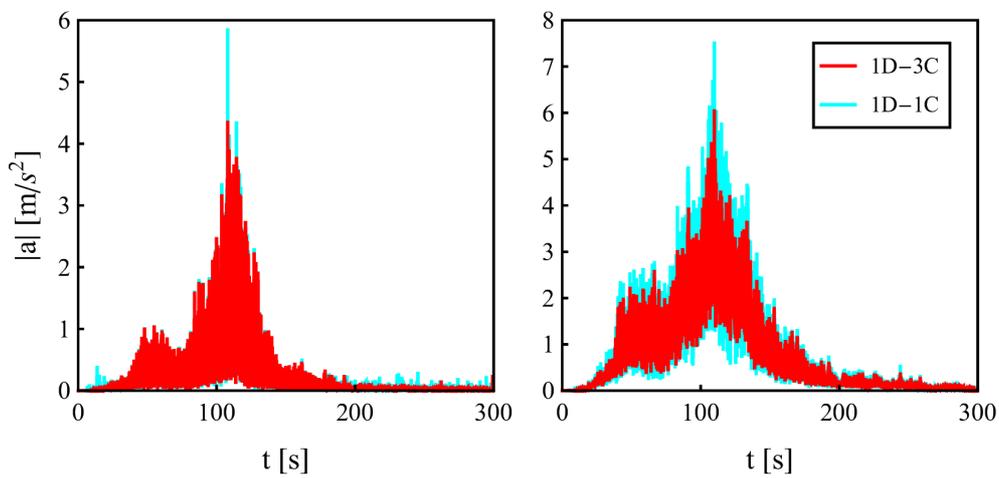

963

964 **Figure 10.** Time history of acceleration modulus at the ground surface during Tohoku

965 earthquake: 1D-3C and 1D-1C numerical solutions for cases FKS011/FKS015 (a) and

966 IBR007/FKS031 (b).

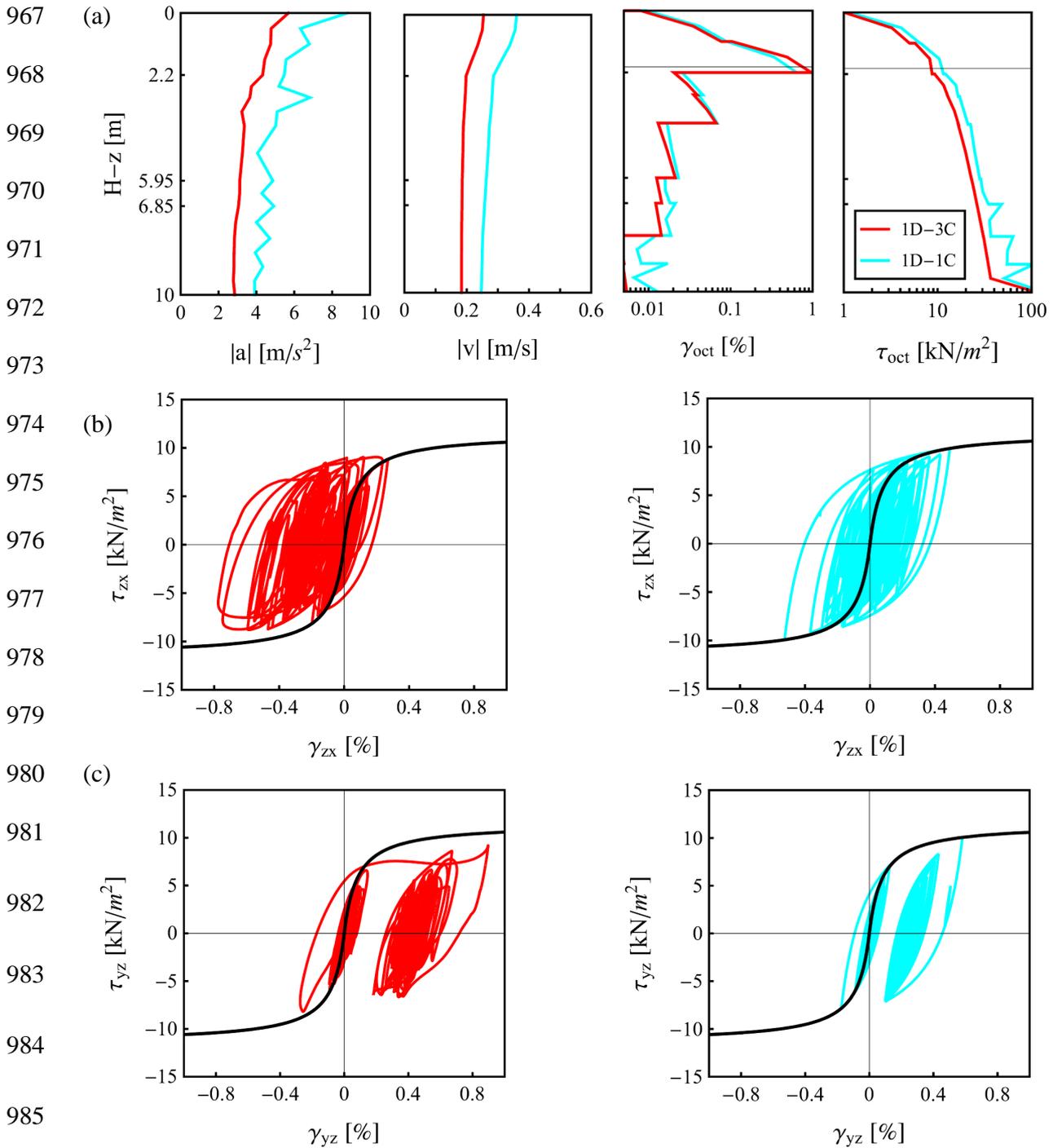

**Figure 11.** 1D-3C and 1D-1C seismic response during the Tohoku earthquake, for the case FKS011/FKS015: profiles of maximum acceleration and velocity modulus, octahedral strain and stress with depth (a); shear stress-strain loops at 2 m depth in x- (b) and y-direction (c).

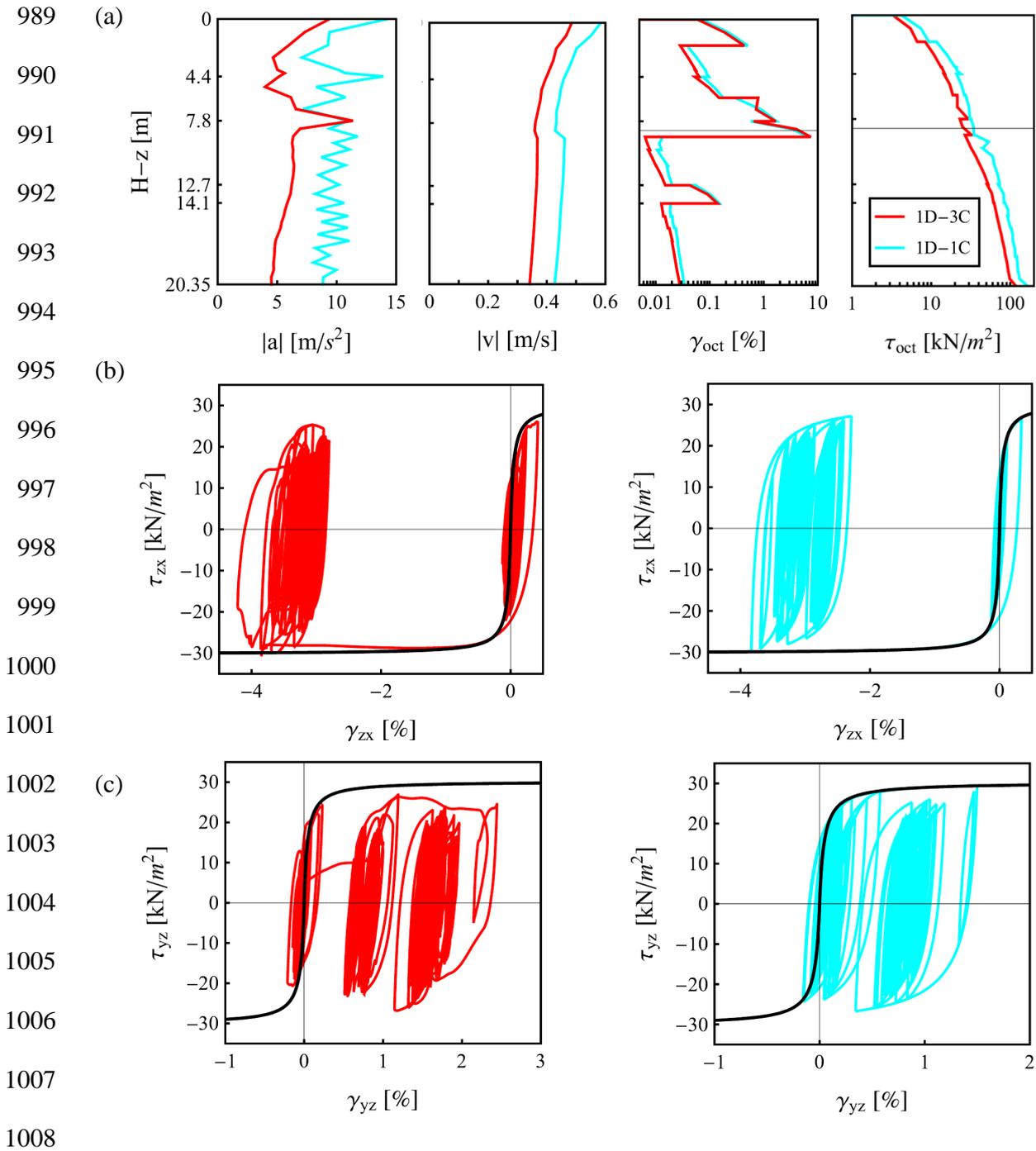

**Figure 12.** 1D-3C and 1D-1C seismic response during the Tohoku earthquake, for the case IBR007/FKS031: profiles of maximum acceleration and velocity modulus, octahedral strain and stress with depth (a); shear stress-strain loops at 8.5 m depth in x- (b) and y-direction (c).

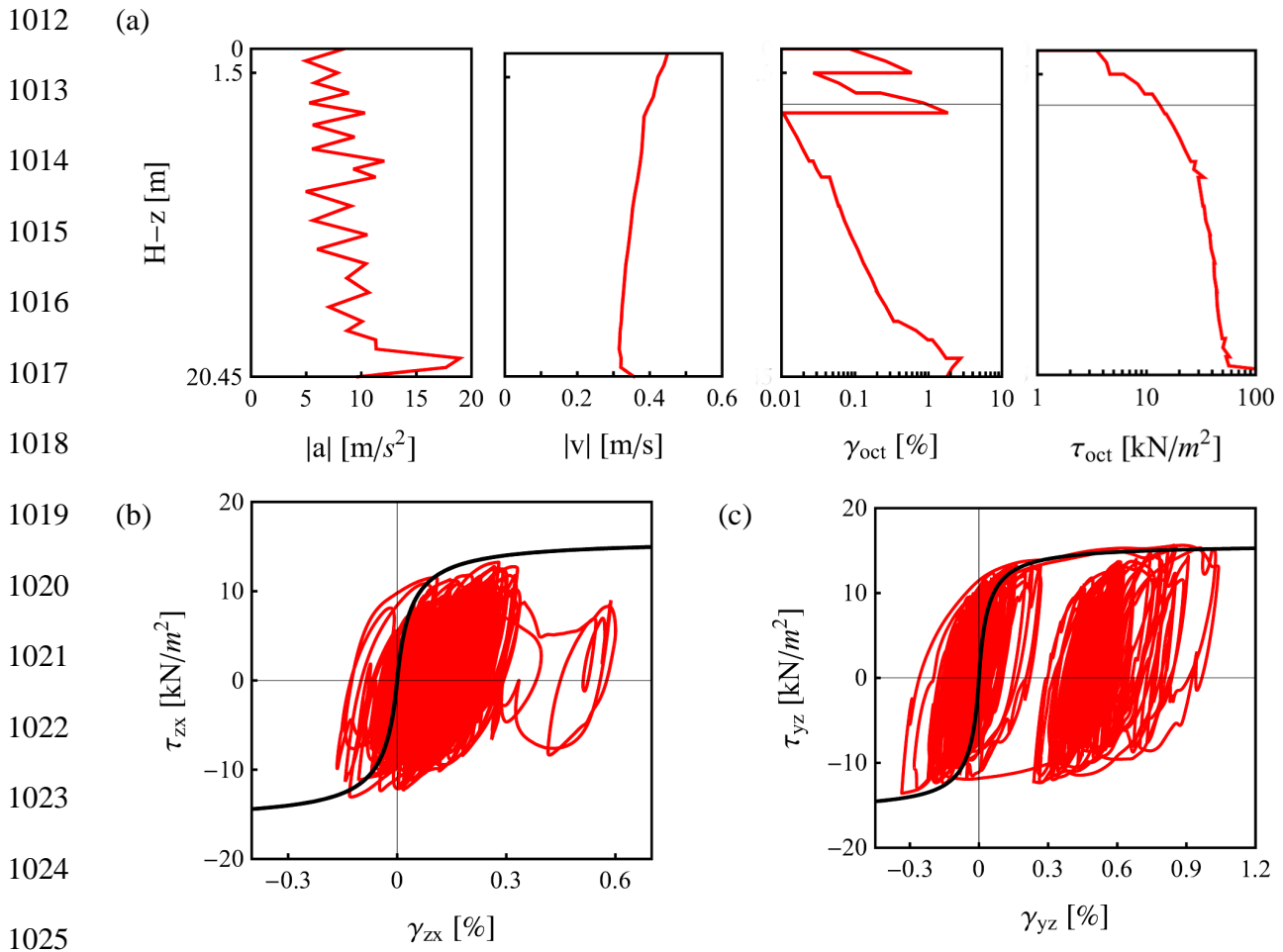

**Figure 13.** 1D-3C and 1D-1C seismic response during the Tohoku earthquake, for the case MYG010/MYG011: profiles of maximum acceleration and velocity modulus, octahedral strain and stress with depth (a); shear stress-strain loops at 3.5 m depth in x- (b) and y-direction (c).

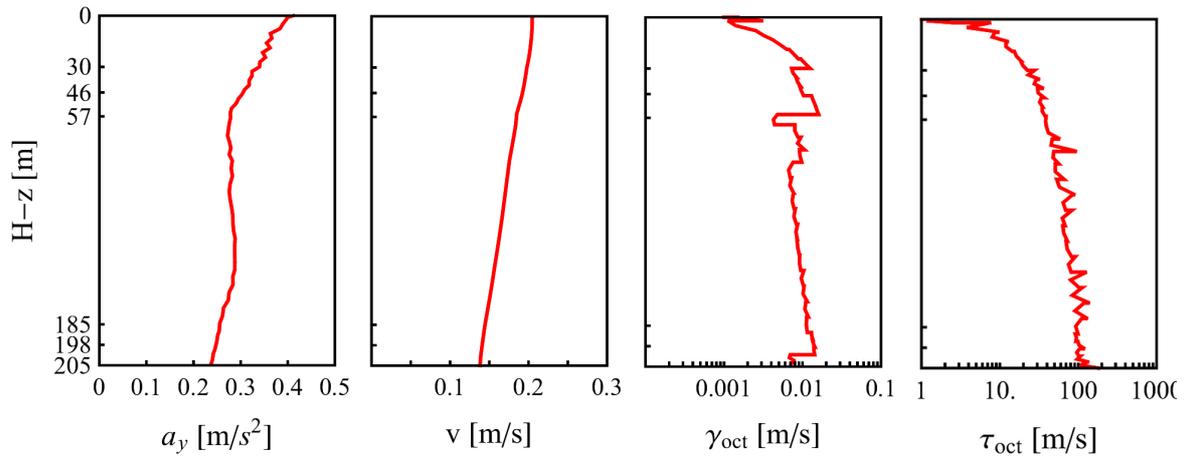

**Figure 14.** Maximum acceleration, velocity, octahedral strain and stress profiles with depth in soil profile NIGH11 during 2011 Tohoku earthquake.

(a) (b)

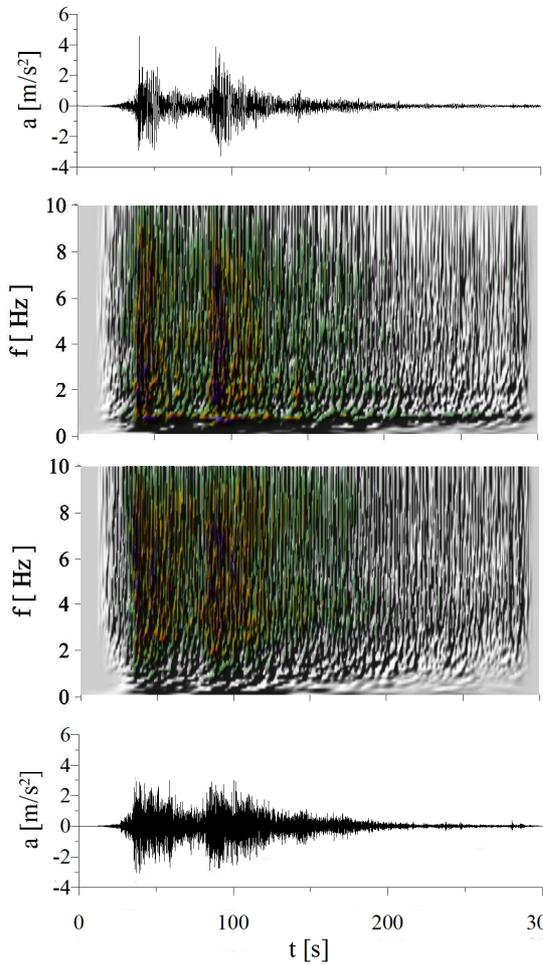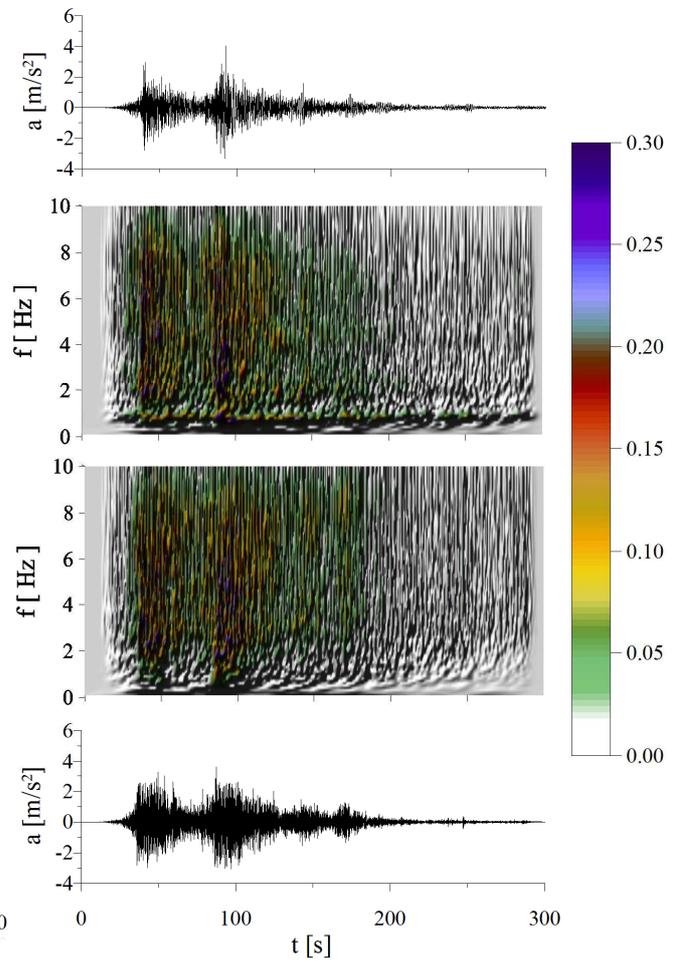

**Figure 15.** Spectral amplitude variation with time and frequency at the ground surface, in horizontal directions x (a) and y (b), during the Tohoku earthquake, evaluated using measured acceleration (up) and computed acceleration (down) as input, for the case MYG010/MYG011.